\documentclass[bibyear]{aa}

\usepackage{graphicx}	
\usepackage[varg]{txfonts}
\usepackage{natbib}
\bibpunct{(}{)}{;}{a}{}{,} 

\usepackage{booktabs}	

\authorrunning{Wutschik et al.}

\title{Star formation and accretion in the circumnuclear disks of active galaxies}
\author{
	Stephanie~Wutschik \inst{\ref{inst1}}
	\and Dominik~R.~G.~Schleicher \inst{\ref{inst1}} 
	\and Thomas~S.~Palmer~II \inst{\ref{inst3}} 
}
\institute{
	Institut f\"ur Astrophysik, Georg-August Universität G\"ottingen, Friedrich-Hund-Platz 1, 37077 G\"ottingen, Germany \\ \email{wutschik@astro.physik.uni-goettingen.de} \label{inst1}
	\and Astronomy Department, Steward Observatory at the University of Arizona, Tucson, AZ \label{inst3} 
}

\date{Received date / Accepted date}

\abstract
{}
{We explore the evolution of supermassive black holes (SMBH) centered in a circumnuclear disk (CND) as a function of the mass supply from the host galaxy and considering different star formation laws, which may give rise to a self-regulation via the injection of supernova-driven turbulence.
}
{A system of equations describing star formation, black hole accretion and angular momentum transport in the disk was solved self-consistently for an axisymmetric disk in which the gravitational potential includes contributions from the black hole, the disk and the hosting galaxy. { Our model extends the framework provided by \citet{Kawakatu2008} by separately considering the inner and outer part of the disk, and by introducing a potentially non-linear dependence of the star formation rate on the gas surface density and the turbulent velocity. The} star formation recipes are calibrated using observational data for NGC~1097, while the accretion model is based on turbulent viscosity as a source of angular momentum transport in a thin viscous accretion disk.}
{We find that current data provide no strong constraint on the star formation recipe, and can in particular not distinguish between models entirely regulated by the surface density, and models including a dependence on the turbulent velocity. The evolution of the black hole mass, on the other hand, strongly depends on the applied star formation law, as well as the mass supply from the host galaxy. We suggest to explore the star formation process in local AGN with high-resolution ALMA observations to break the degeneracy between different star formation models.}
{}


\begin{document}
\maketitle

%

\section{Introduction}			\label{sec:introduction}

Supermassive black holes are observed in the centers of virtually all galaxies, and their properties are tightly correlated to the mass of the stellar bulge and its velocity dispersion \citep{Magorrian98, Ferrarese00, Gebhardt00, Graham01, Merritt01, Tremaine02, Haering04}. The correlation between the stellar mass and its velocity dispersion indeed suggests a link between the evolution of galaxies and their central black hole, or between the star formation rates and the black hole accretion rate. While the presence of supermassive black holes in the local Universe can be understood in terms of Eddington accretion, the latter appears much more difficult for supermassive black holes at $z\gtrsim6$ \citep{Shapiro05}.

Black holes with more than $10^9$~M$_\sun$ have however been confirmed even beyond $z=6$, both with the Sloan Digital Sky Survey (SDSS) \citep{Fan2001, Fan03, Fan04, Fan06, Fan06b},  the Canada-France high-redshift Quasar survey \citep{Willott07} or the UKIDSS Large Area Survey \citep{Venemans07}. The currently highest-redshift detection corresponds to a $2\times10^9$~M$_\sun$ black hole at $z=7.085$ \citep{Mortlock2011}, about $760$~million~years after the Big Bang. The formation of supermassive black holes thus likely requires a mechanism of efficient accretion and angular momentum transport. 

While high-redshift black holes are particularly challenging from a theoretical perspective, active galactic nuclei (AGN) in our local environment provide a relevant test case in order to probe the mechanisms for accretion and angular momentum transport in realistic systems. Observations reported the occurrence of circumnuclear disks and starburst rings in many of these systems. NGC~1068, the prototype AGN in our local neighborhood, has been explored in the stellar light \citep{Telesco84, Davies07}, but also in molecular gas and dust continuum by \citet{Schinnerer2000, Galliano2003, Galliano05}, finding a $3$~kpc-scale ring of molecular gas with intense star formation activity. 

In a similar manner, NGC~1097 shows a molecular ring combined with intense star formation activity on scales of $\sim700$~pc \citep{Hummel87, Telesco81, Kotilainen00}, and the presence of dense gas has been inferred via CO and HCN observations by \citet{Kohno03} and \citet{Hsieh08}. The latter indicate typical densities of $10^3$~cm$^{-3}$ and temperatures of $\sim100$~K, comparable to the conditions in  starburst galaxies \citep{Wild92, Aalto95, Loenen10}. Indeed, more recent studies by \citet{Hsieh2011} have for the first time resolved individual giant molecular cloud complexes, allowing to test the correlation between molecular gas and star formation on more local scales. With ALMA\footnote{ALMA telescope: http://www.almaobservatory.org/}, accretion in these galaxies has now been probed down to scales of $40$~pc \citep{Fathi13}. Radio observations suggest the presence of strong magnetic fields with up to $60$~$\mu$G, which may thus contribute to the accretion process \citep{Beck99, Beck05}.

Indeed, circumncuclear disks and rings appear to be ubiquitos in local AGN, and can thus be expected to play a fundamental role for black hole accretion \citep{Davies07}. A systematic study concerning the correlation between the black hole accretion rate and the star formation rate has been pursued by \citet{Stanic12}. While such a correlation appears to exist on all scales, it appears particularly strong below $1$~kpc, i.e. on scales corresponding to the circumnuclear disk, where they report a scaling relation of SFR $\propto \dot{M}_\mathrm{BH}^{0.8}$. A strong connection between star formation and black hole accretion has further been indicated by Herschel\footnote{Herschel satellite: http://sci.esa.int/science-e/www/area/index.cfm?fareaid=16} observations of Mrk~231 \citep{Werf10}, as an analysis of the high-$J$ CO lines provides indications both for strong photon-dominated regions (PDRs) and X-ray dominated regions (XDRs). The latter does correspond to the feedback from stars and a supermassive black hole. While Herschel was probing this phenomenon for nearby AGN, we note here that ALMA may provide the prespective of detecting X-ray dominated regions even at high redshift \citep{Schleicher10c}.

While observational probes become more difficult at high redshift, the presence of dust and molecular gas in $z>4$ quasars has nevertheless been reported already by \citet{Omont96a, Omont96b, Carilli02}. The first detections in a $z=6.42$ quasar have been pursued by \citet{Walter04, Riechers07, Walter09}, revealing the presence of a molecular gas mass of $4.5\times10^{10}$~M$_\sun$. Even for the highest-redshift black hole at $z=7.085$, dust and [CII] emission have been reported by \citet{Venemans12}.

Particularly well-suited for the study of the circumnuclear accretion disks in high-redshift quasars are lensed systems, which reveal the dynamics in the central $100-500$~pc. An especially well-studied system is APM~08279+5255 at $z=3.9$, where CO and HCN observations indicate warm molecular gas with densities of $\sim10^5$~cm$^{-3}$ \citep{Weiss07, Riechers10}, while detections of water lines indicate an Eddington-limited starburst on similar scales \citep{Werf11}. Similar conditions have been found for a lensed quasar at $z=4.1$ \citep{Riechers08} and the Cloverleaf quasar at $z=2.56$, where warm molecular gas was detected both in CO \citep{Bradford09} and HCN \citep{Riechers11}.

Based on these even though limited observations, we may thus conclude that circumnuclear starburst rings may play a significant role in connecting black hole growth and star formation both at low and high redshift. Such a connection has indeed been suggested also in theoretical models \citep{Thompson05, Levin07, Kawakatu2008, Vollmer08, Kawakatu09, Kumar2010}, as stellar feedback may drive strong supersonic turbulence, which may drive the accretion via turbulent viscosity \citep[e.g.][]{Shakura1973, Duschl11}. Indeed, enhanced accretion has been observed in numerical simulations in the presence of highly supersonic turbulence \citep{Hobbs11}, and more realistic approaches aim to self-consistently inject turbulent energy via supernova-explosions \citep{Wada09}. Other attempts have focused on the impact of black hole feedback on nearby star-forming clouds \citep{Hocuk11}, including the potential effects of magnetic fields \citep{Hocuk12}.

While it is known that turbulence injected via stellar feedback may enhance the accretion, numerical simulations allow to explore only a limited range of conditions, while previous semi-analytic models required simplifying assumptions concerning the structure of the disk. As high-resolution observations with ALMA will allow to probe the structure of these disks, we aim here to extend the { model of \citet{Kawakatu2008, Kawakatu09} by including a more detailed description of the interior structure, and by considering a potentially nonlinear dependence of the star formation rate on the gas surface density and the turbulent velocity}. Of course, such an attempt still requires assumptions to be made, which can however be validated both with upcoming observations as well as numerical simulations.  In section 2, we provide the outline of our model, which is then applied in section 3 to different star formation models. We report here the impact of different star formation descriptions, as the latter are still uncertain both from a theoretical and an observational point of view. A discussion of these results is then provided in section 4, and the main conclusions are summarized in section 5. 
\section{Outline of the model}		\label{sec:model}

\begin{figure}[t] 
 \centering
 \includegraphics[width=8.8 cm]{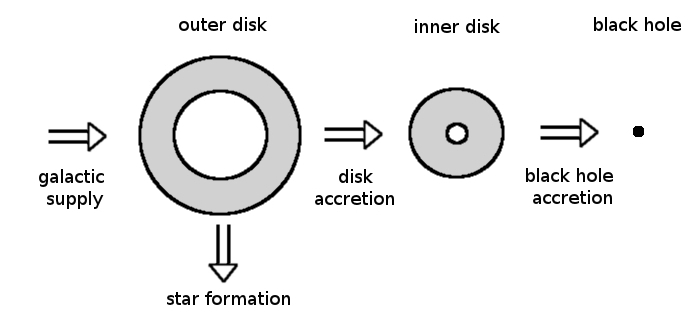}
 \caption{Diagramm of the mass flow between the inner and the outer gaseous disk. Matter is supplied from the hosting galaxy to the outer nuclear disk, which forms stars, thereby reducing the amount of available gas. Via accretion gas can flow from the outer disk into the inner disk, which is gravitationally stable. The inner gas disk is depleted by the accreting black hole in the center.}
 \label{fig:massflow}
\end{figure}

The concept of this model is based on a self-gravitating gaseous disk around a SMBH. We supply the system with dusty gas at a time-dependent rate $\dot{M}_\mathrm{sup}$ from a host galaxy with a surface density $\Sigma_\mathrm{host}$, including gas and stars, 
where the dust to gas ratio is similar to the solar neighbourhood.
The accumulating gas and dust grains form a disk around the central seed black hole with mass $M_\mathrm{BH}$ as described in Fig. \ref{fig:massflow}.
All components are rotating around the central black hole with an angular velocity $\Omega(r) = G M_\mathrm{BH} / r^3 + G \pi / r ( \Sigma_\mathrm{disk} + \Sigma_\mathrm{host} )$.

In this calculation, our model assumes that the disks are dominated by the gas component, which we expect to be a good assumption in the active accretion phase and during intense starbursts, requiring ubiquitous molecular gas for intense star formation activity. This assumption becomes less accurate at late times when the stellar mass becomes comparable and the molecular gas reservoir is exhausted. We are however not predominantly interested in these phases, and we have checked that the resulting mass ratios remain essentially unchanged if we account for the additional stellar mass in the angular velocity. 

We assume an isothermal disk with a constant sound speed. We take into account that only part of the disk might be gravitationally unstable, separately considering the evolution of the stable part of the disk (inner disk) and the self-gravitating part (outer diks). We apply different star formation models and consider the amount of turbulence introduced into the gas by supernovae, which strongly influences the accretion onto the black hole. For the accretion process we adopt the formula for a viscous accretion disk as formulated by \citet[][see \citet{Lodato2008} for a review]{Pringle1981}. 

\subsection{Radial disk structure}		\label{ssec:diskrad}
The outer radius of the disk is defined by $r_\mathrm{out}$ given in Eq. (\ref{eq:rout}) and marks the farthest reach of the dominion of the gravitational potential of the system with respect to the host galaxy. If the disk mass $ M_\mathrm{disk} = \int_{r_\mathrm{in}}^{r_\mathrm{out}} 2 \pi r' \Sigma_\mathrm{disk}(r') dr' $ with the inner radius of the disk $ r_\mathrm{in} $ dominates the gravitational potential inside the disk, we obtain $ r_\mathrm{out} $ from $ G M_\mathrm{disk} / r_\mathrm{out}  =  \pi G ~ \Sigma_\mathrm{host} ~ r_\mathrm{out} $. On the other hand, if the gravitational potential inside the disk is dominated by the black hole, we obtain $ r_\mathrm{out} $ from $ G M_\mathrm{BH} / r_\mathrm{out} = \pi G ~ \Sigma_\mathrm{host} ~ r_\mathrm{out} $.
\begin{equation}
	r_\mathrm{out}= \left\lbrace
						\begin{array}{ll}
                        \sqrt{\frac{M_\mathrm{disk}}{\pi \Sigma_\mathrm{host}}}, \ & M_\mathrm{disk} > M_\mathrm{BH}\\
                        \sqrt{\frac{M_\mathrm{BH}}{\pi \Sigma_\mathrm{host}}},\  & \mathrm{else}
                       \end{array}\right.
	\label{eq:rout}
\end{equation}
There $M_\mathrm{disk} = M_\mathrm{gas}+M_*$ is the sum of the gaseous mass $M_\mathrm{gas}$ and the stellar mass $M_*$.

The inner radius of the disk is defined by $r_\mathrm{in}$ which is assumed to be determined by the sublimation radius of silicon dust. This assumption is supported by observational results \citep{Suganuma2006}. If the AGN luminosity $L_\mathrm{AGN}$ heating up the dust grains to the sublimation temperature of 1500 K, equals the Eddington luminosity $L_\mathrm{Edd}=4 \pi c \ G M_\mathrm{BH}\ m_\mathrm{p}/\sigma_\mathrm{T}$ with the proton mass $m_\mathrm{p}$ and the Thomson cross section $\sigma_\mathrm{T}$, the inner radius can be computed to $r_\mathrm{in}=3 {\mathrm \ pc \ } \sqrt{M_\mathrm{BH}\ \times 10^{-8}}$. We note that the calculated inner radius is a maximum radius. 
Especially during slow accretion phases the actual inner radius might be closer to the central black hole. 

As gas is supplied to the system the disk will become gravitationally unstable eventually, giving rise to star formation which will trigger supernova feedback. To identify the region of instability we use the Toomre-function $Q(r)$ to calculate the radius, at which the disk is just stable, i.e. at which $Q(r) = 1$.
This we call the critical radius $r_\mathrm{c}$ so that the disk is self-gravitating at radii $r > r_\mathrm{c}$, which essentially determines the region of star formation and supernova feedback.
The Toomre function is determined as the fraction of the critical surface density $\Sigma_\mathrm{crit}(r)$ over the gas surface density $\Sigma_\mathrm{gas}(r)$ where $\Sigma_\mathrm{crit}$ is determined via
\begin{equation}
	\Sigma_\mathrm{crit} (r) = \frac{\kappa(r) c_\mathrm{s}}{\pi G}.
	\label{eq:critsurf}
\end{equation}
There, $\kappa^2 (r) = 4\Omega^2(r) + 2 \Omega(r) \cdot r \mathrm{d}\Omega(r)/\mathrm{d}r$ is the epicyclic frequency and $c_\mathrm{s}$ is the sound speed.
We find the critical radius $r_\mathrm{c}$ at $Q(r) = \Sigma_\mathrm{crit}(r) / \Sigma_\mathrm{gas}(r) = 1$, where the surface density of the gas is equal to the critical surface density.
The gas surface density is calculated  for the inner and the outer disk:
\begin{eqnarray}
	\Sigma_\mathrm{out}(r)&=& \Sigma_\mathrm{g}(r\geq r_\mathrm{c})=\Sigma_0\cdot (r/r_\mathrm{c})^{-\gamma}\label{eq:surfout}\\
	\Sigma_\mathrm{in}(r)&=& \Sigma_\mathrm{g}(r < r_\mathrm{c})=\Sigma_0\cdot (r/r_\mathrm{c})^{-\gamma^*}\label{eq:surfin}
\end{eqnarray}
where $\Sigma_0$ is computed from $M_\mathrm{g,o}=2\pi \int_{r_\mathrm{c}}^{r_\mathrm{out}} r \,\Sigma_\mathrm{out}(r) \,  \mathrm{d}r $ and $\gamma$ is a free parameter. For the inner disk $\gamma^*$ is obtained from \mbox{$M_\mathrm{g,i} = 2 \pi \int_{r_\mathrm{in}}^{r_\mathrm{c}} r \,\Sigma_\mathrm{in}(r) \, \mathrm{d}r $}. In case of a totally stable disk, i.e. no outer disk exists, $\Sigma_0$ is calculated from \mbox{$M_\mathrm{g,i} = 2 \pi \int_{r_{\mathrm{in}}}^{r_\mathrm{out}} r \Sigma_0 \ (r/r_\mathrm{c})^{-\gamma} \mathrm{d}r$}.

\subsection{Vertical structure}	\label{ssec:diskvert}	
We have two regimes which determine the vertical structure, i.e. the scaleheight of the disk.

The first regime is the subsonic turbulent regime, where the thermal pressure is greater than the turbulent pressure and therefore determines the scaleheight.
We assume hydrodynamical equilibrium so that $\Sigma_\mathrm{gas}(r) \cdot c_\mathrm{s}^2 = \Sigma_\mathrm{gas}(r) \cdot g \cdot h_\mathrm{therm}(r)$ where $g = G M_\mathrm{BH} \cdot h(r) / r^3 + \pi \left( \Sigma_\mathrm{gas}(r) + \Sigma_\mathrm{host} \right)$ is the local gravity.
There the second term can be neglected for smaller radii, where the gravitational potential is dominated by the central black hole. In this case, we obtain expression (\ref{eq:htherm}) for the thermal scaleheight
\begin{equation}
	\label{eq:htherm}
	h_\mathrm{therm}(r) = \sqrt{\frac{r^3}{G~M_\mathrm{BH}}} c_\mathrm{s}
\end{equation}
and the turbulent velocity is equal to the sound velocity throughout the subsonic turbulent region.

The second regime is the supersonic turbulent regime, where the turbulent pressure determines the scaleheight. 
Observations indicate, that active galactic nuclei often show violent star formation in the galactic nucleus as well, 
e.g. \citet{LaMassa2013, Santini2012}. 
An intense star formation can trigger feedback mechanisms, which can deposit significant amounts of energy in the surrounding ISM. In this model we consider supernovae, which will deposit part of their thermal energy into the ISM as kinetic energy.
Due to the vertical hydrostatic balance \citep{Shetty12}, $E_\mathrm{turb}(r) = \Sigma_\mathrm{gas}(r) \cdot g \cdot h(r)$ with $E_\mathrm{turb}(r)$ the turbulent energy per unit area at radius $r$, we obtain the turbulent scaleheight from equation (\ref{eq:hturb}).

\begin{equation}
	\label{eq:hturb}
	h_\mathrm{turb}(r) = \sqrt{\frac{r^3}{G~M_\mathrm{BH}} \frac{E_\mathrm{turb}(r)}{\Sigma_\mathrm{gas}(r)}}
\end{equation}

On the other hand, the turbulent velocity $v_\mathrm{turb}(r)$ is obtained directly from $E_\mathrm{turb}(r) = \Sigma_\mathrm{gas}(r)\cdot v_\mathrm{turb}^2(r)$.

In our model, the evolution of the turbulent energy is described by the differential equation
\begin{equation}
	\label{eq:eturb-evolution}
	\frac{\mathrm{d}E_\mathrm{turb}(r)}{\mathrm{d}t} = \dot{E}_\mathrm{inj}(r) - \dot{E}_\mathrm{dis}(r)
\end{equation}
where $ \dot{E}_\mathrm{inj}(r) $ is the rate at which energy is injected into the medium and $ \dot{E}_\mathrm{dis}(r) = E_\mathrm{turb}(r) / t_\mathrm{dis}(r)$ is the energy dissipation rate with the dissipation time scale $ t_\mathrm{dis} = h_\mathrm{turb}(r) / v_\mathrm{turb}(r) $. This implies, that the dissipation timescale is equal to a crossing time. This equation is solved numerically via the common fourth-order Runge-Kutta method for each time $t$.

For the energy-injection we calculate the energy input via supernova explosions and obtain
\begin{equation}
	\label{eq:snfb_1}
	\dot{E}_\mathrm{inj}(r,t) = f_\mathrm{SN} \eta_\mathrm{SN} E_\mathrm{SN} \cdot \hat{\dot{\xi}}_*(r,t - T_\mathrm{SN})
\end{equation}
where $ \hat{\dot{\xi}}_*(r, t - T_\mathrm{SN}) $ is the local star formation rate per unit area at time $t-T_{\mathrm{SN}}$ with $T_\mathrm{SN} \approx 10^6 \mathrm{yr}$ being the average life time of massive stars, that explode into core-collaps supernovae, i.e. stars with initial masses greater than eight solar masses\footnote{obtained from the IMF where the most common IMFs assume the same exponents for the high-mass regime, see also appendix \ref{appendix:imf}}. \mbox{$ f_\mathrm{SN}= 7.9  \times 10^{-3} $} is the supernova rate per solar mass of formed stars\footnotemark[\value{footnote}], $ \eta_\mathrm{SN} $ is a heating efficiency and $ E_\mathrm{SN} = 10^{51} \mathrm{erg}$ is the thermal energy typically injected by core-collapse supernovae.

For easier calculation of the local supernova rate, we use a modified star formation rate surface density $ \tilde{\dot{\xi}} _*(r, t- T_\mathrm{SN}) = A ~ (r / r_\mathrm{c} )^{ -\theta \gamma - \lambda \epsilon } $ where the integration of \mbox{$ \tilde{\dot{\xi}} _*(r, t- T_\mathrm{SN}) $} over the whole disk yields the absolute star formation rate $ \hat{ \dot{M} }_* (t-T_\mathrm{SN}) $ and
\begin{equation}
A = \frac{ \hat{ \dot{M} }_* }{2 \pi} (2 - \theta \gamma - \epsilon \lambda) r_\mathrm{c}^{ - \theta \gamma - \epsilon \lambda } \left[ r_\mathrm{out}^{ 2 - \theta \gamma - \epsilon \lambda } - r_\mathrm{in}^{ 2 - \theta \gamma - \epsilon \lambda } \right]^{-1}
\end{equation}
where $ \hat{ \dot{M} }_* (t-T_\mathrm{SN}) $ is obtained from the integration of the original star formation rate per unit area over the currently star forming region, as the typical disk properties are expected to evolve over longer timescales than a typical supernova explosion. The  local energy injection rate per unit area is then given as\begin{eqnarray}
	\dot{E}_\mathrm{inj}(r) &=& 7.9 \times 10^{-3} \ \eta_\mathrm{SN} E_\mathrm{SN} \ \frac{\hat{\dot{M}}_*}{2 \pi} \ ( 2 - \theta \gamma - \epsilon \lambda ) \cdot r^{ - \theta \gamma - \epsilon \lambda } 	\nonumber\\
	&&\times \left[ r_\mathrm{out}^{ 2 - \theta \gamma - \epsilon \lambda } - r_\mathrm{in}^{ 2 - \theta \gamma - \epsilon \lambda } \right]^{-1}.	\label{eq:snfb_2} 
\end{eqnarray}

\subsection{Star formation}		\label{ssec:sfr}

The star formation rate in the disk $\dot{M}_*$ can be calculated by integrating the local star formation rate per unit area over the whole star forming region: $\int_{r_\mathrm{min}}^{r_\mathrm{out}} 2\pi ~ r \cdot \dot{\xi}_* (r) ~ \mathrm{d}r$ where $r_\mathrm{min}$  is either the inner radius $ r_\mathrm{in} $ ($r_\mathrm{c} \leq r_\mathrm{in}$) or the critical radius $r_\mathrm{c}$ ($r_\mathrm{c} > r_\mathrm{in}$).
As we want to study the influence of the star formation model on the evolution of the system, we parametrize the star formation rate per unit area as follows:
\begin{equation}
	\dot{\xi}_*(r) = \Psi \cdot \left( \Sigma_\mathrm{gas}(r) \right)^\theta \cdot (v_\mathrm{turb}(r))^{- \epsilon}
	\label{eq:sfr_rad}
\end{equation}
There $\Psi$ is the normalization constant and $ 0 \leq \theta \leq 2 $ and $ 0 \leq \epsilon \leq 1$ are free parameters.

\begin{table}[tbh]
	\caption{Star formation model parameters}
	\begin{tabular}{ccc ccc}\hline\hline\addlinespace[0.1cm]
	$\epsilon = 0$	&	&				&	$\epsilon = 1$	&		&			\\
	model	&	$\theta$	&	$\Psi$	&	model	&	$\theta$	&	$\Psi$	\\\hline\addlinespace[0.1cm]
	U0		&	0.5		&	$10^{-6}$	&	&&\\
	U1		&	1.0		&	$10^{-8}$	&	S1		&	1.0		&	$10^{-11}$	\\
	U2		&	1.25	&	$10^{-9}$	&	S2		&	1.25	&	$10^{-12}$	\\
	U3		&	1.5		&	$10^{-10}$	&	S3		&	1.5		&	$10^{-13}$	\\
	U4		&	1.75	&	$10^{-11}$	&	S4		&	2.0		&	$5\times10^{-16}$	\\\hline
	\end{tabular}
\label{tab:parameters_sfr}
\end{table}

This allows us to apply a great diversity of both velocity-independent ($\epsilon = 0$) and self-regulating ($\epsilon = 1$) star formation models e.g. the models suggested by \citet{Kawakatu2008} and \citet{Elmegreen2010}. Table \ref{tab:parameters_sfr} lists the star formation models we applied in this work, where U1 rouhly corresponds to the model studied by \citet{Kawakatu2008} and S4 corresponds to the model proposed by \citet{Elmegreen2010}.

To obtain the total star formation rate we need to calculate the integral of $ \dot{\xi}_*(r) $ from the critical radius $r_\mathrm{c}$ to the outer radius of the disk $r_\mathrm{out}$. In order to carry out this integration, we approximate the turbulent velocity as a power law in Eq. (\ref{eq:vpower}). We obtain

\begin{eqnarray}
v_\mathrm{turb}(r) &=& v_0 \cdot \left( \frac{r}{r_\mathrm{out}} \right)^\lambda		\label{eq:vpower}\\
\textrm{ with } v_0 &=& \sqrt{E_\mathrm{turb}(r_\mathrm{c}) / \Sigma_\mathrm{gas}(r_\mathrm{c})}\\
\textrm{ and } \lambda &=& \frac{1.5 - \gamma (\theta - 1)}{2 + \epsilon},
\end{eqnarray}

where we choose $v_0 = c_\mathrm{s}$ and $\lambda = 0$ in case of (sub-)sonic turbulence.
This leads to eq. (\ref{eq:sfr_tot}), which gives the star formation rate at time $t$ integrated over the disk.

\begin{equation}
	\label{eq:sfr_tot}
\dot{M}_*(t) = \frac{ 2 \pi \cdot \Psi }{ 2 - \theta \gamma - \epsilon \lambda } \cdot \Sigma_0^\theta \cdot v_0^\epsilon r_\mathrm{c}^{\theta \gamma + \epsilon \lambda} \left( r_\mathrm{out}^{2 - \theta \gamma - \epsilon \lambda} - r_\mathrm{c}^{2 - \theta \gamma - \epsilon \lambda} \right)
\end{equation}

The free parameters $\theta$, $\gamma$ and $\epsilon$ have to be chosen carefully to ensure that $\theta \gamma + \epsilon \lambda \neq 2$.

\subsection{Black hole accretion} \label{ssec:bh-acc}
In order to accrete matter from the circumnuclear disk onto the central black hole angular momentum has to be transported by some mechanism. In this model we will follow the $\alpha$-viscosity-prescription by \citet{Shakura1973} in the approximation for a thin disk, as formulated by \citet{Pringle1981}. Equation (\ref{eq:accretion}) gives the expression for accreted matter at radius $R$, as given by \citet{Kawakatu2008}.
\begin{equation}
\dot{M}(R,t)= 2\pi\nu(r)\ \Sigma(R,t) \left|\frac{d\ln\Omega(R,t)}{d\ln R}\right|
\label{eq:accretion}
\end{equation}
There, the viscosity $ \nu(r) = \alpha \ v(r) \cdot h(r) $ is i) determined by supersonic turbulence in case of a self-gravitating disk or ii) caused by magneto-rotational instabilities, resulting into subsonic turbulence. In the first case, $ \alpha $ is of the order of unity, $ v(r) $ is the turbulent velocity $ v_\mathrm{turb}(r) $ and $ h(r) $ is the turbulent scaleheight $ h_\mathrm{turb}(r) $, determined by the turbulent pressure caused by stellar feedback. In the second case, the scaleheight is dominated by thermal pressure. 
The turbulent velocities are comparable to sound velocity or smaller, resulting into $ \alpha \approx 0.01 - 0.5 $ \citep[see][and references therein]{Kawakatu2008}. 
To calculate the accretion rate of the black hole, we evaluate Eq. (\ref{eq:accretion}) at the inner radius of the disk $ r_\mathrm{in} $, assuming that all matter crossing that radius will fall into the black hole eventually.
\subsection{Matter transport inside the disk}

In general, we assume that the matter infall from the host galaxy is significantly less efficient than the redistribution inside the disk. Therefore we assume, that the exponent of the gas surface density power law is not influenced by the infalling matter.

As we consider the inner and outer disk seperately with individual power law surface densities, we need to take into account the transport of the gas from the outer to the inner disk and vice versa. The matter exchange consists of two contributions: \emph{i)} a physical transport of matter from the outer to the inner disk due to accretion processes and \emph{ii)} a purely geometrical transport process due to the evolution of the critical radius.

\emph{i)} If the disk is only partially self-gravitating, we assume, that the surface density in the inner and outer disk obey power laws with different exponents (see sect. \ref{ssec:diskrad}). As the disk is supplied with additional gas from the host galaxy, this matter flows into the outer disk. However, the inner part of the disk is able to accrete matter from the outer disk with the same transport mechanism as the black hole accretes matter from the inner disk. We take this matter transport into account by calculating the disk accretion rate at times $ t $, evaluating Eq. (\ref{eq:accretion}) at \mbox{the critical radius $ r_\mathrm{c} $}.

\emph{ii)} If matter in the outer disk, which is gravitationally unstable, becomes stable, the critical radius moves outward. In this case, the now-stable matter has to be moved from the outer gas reservoir to the inner gas reservoir and the gas surface densities of both inner and outer disk have to be recalculated. To determine the amount of gas that has to become part of the inner disk, we calculate the integral of the gas surface density from the former critical radius to the updated critical radius, using the power law of the outer disk. In case of an inward directed critical radius (gravitationally stable matter becomes self-gravitating) the integration is calculated with the surface density power law of the inner disk and the matter is moved from the inner gas mass to the outer gas mass.
The general formulation
\begin{equation}
	\label{eq:disksup}
\Delta M_\mathrm{gas} = 2 \pi \int_{r_\mathrm{c}(t_0)}^{r_\mathrm{c}(t)} r \cdot \Sigma_0(t_0) \left( \frac{r}{r_\mathrm{c}(t_0)} \right)^\beta \mathrm{d}r' \; ,
\end{equation}
where $\beta = \gamma$ if $r_\mathrm{c}(t_0) < r_\mathrm{c}(t)$ and $\beta = \gamma^*$ if $r_\mathrm{c}(t_0) > r_\mathrm{c}(t)$, results into a positive value if gas has become stable ($r_\mathrm{c}(t_0) < r_\mathrm{c}(t)$) and into a negative value, if gas has become self-gravitating ($r_\mathrm{c}(t_0) > r_\mathrm{c}(t)$). So we obtain the updated gas masses $M_\mathrm{g,o}(t) = M_\mathrm{g,o}(t_0) - \Delta M_\mathrm{gas}$ and $M_\mathrm{g,i}(t) = M_\mathrm{g,i}(t_0) + \Delta M_\mathrm{gas}$.

\subsection{Model parameters}

To model the system of the central black hole embedded in a circumnuclear disk we need to specify several free parameters. Table~\ref{tab:parameters_general} gives the parameters which were chosen to correspond to those utilized by \citet{Kawakatu2008} for better comparability.

\begin{table}[tbh]
	\caption{General model parameters}
	\begin{tabular}{cll}\hline\hline\addlinespace[0.1cm]
	parameter	&	description		&	value		\\\hline\addlinespace[0.1cm]
	$\Sigma_\mathrm{host}$	&	galactic gas surface density	&	$10^4 \, M_\sun \,\mathrm{pc^{-2}}$\\
	$p_\mathrm{sup}$	&	galactic mass supply rate	&	$1.0 \, M_\sun \,\mathrm{yr^{-1}}$	\\
	$M_{\mathrm{BH},0}$	&	black hole seed mass	&	$10^3 \,M_\sun$	\\
	$c_s$				&	sound velocity	&	$10^3 \, \mathrm{m \,s^{-1}}$	\\
	$\gamma$			&	gas surf. dens. power-law exp.	&	1.0			\\
	$\alpha$			&	efficiency for subsonic accretion	&	1.0		\\
	$\eta$				&	supernova heating efficiency	&	0.1			\\\hline
	\end{tabular}
	\label{tab:parameters_general}
\end{table}
While the value of $\alpha$ is uncertain in both regimes, we chose $\alpha=1$ in order to have a smooth transition between the two different accretion processes. As the contribution of subsonic accretion to the final black hole mass is negligible with accretion rates two orders of magnitude smaller than supersonic accretion, the specific choice of $\alpha$ in the stable phase of the disk has no impact on our final results.

\section{Results}

\subsection{Dynamical evolution of the system with unregulated star formation}\label{ssec:evo_ur-sfr}

As described in section~\ref{ssec:sfr} we apply different star formation models which are parametrized by the three parameters $\theta$, $\epsilon$ and $\Psi$. For unregulated star formation we have $\epsilon \equiv 0$ so that the star formation model is insensitive to an increased turbulent velocity (compare Eq.~\ref{eq:sfr_rad}). $\theta$ is a free parameter, whereas $\Psi$ has been chosen in such manner that the final stellar masses are roughly equal for all pairs of $(\theta, \Psi)$, to ensure comparability of the different models.
Table~\ref{tab:parameters_sfr} lists the studied models of unregulated star formation where { U1} is equivalent to the model proposed by \citet{Kawakatu2008} who chose $\theta = 1$ and $\Psi = 3 \times 10^{-8}$.

\begin{figure}[bt]
	\centering
	\resizebox{\hsize}{!}{\includegraphics{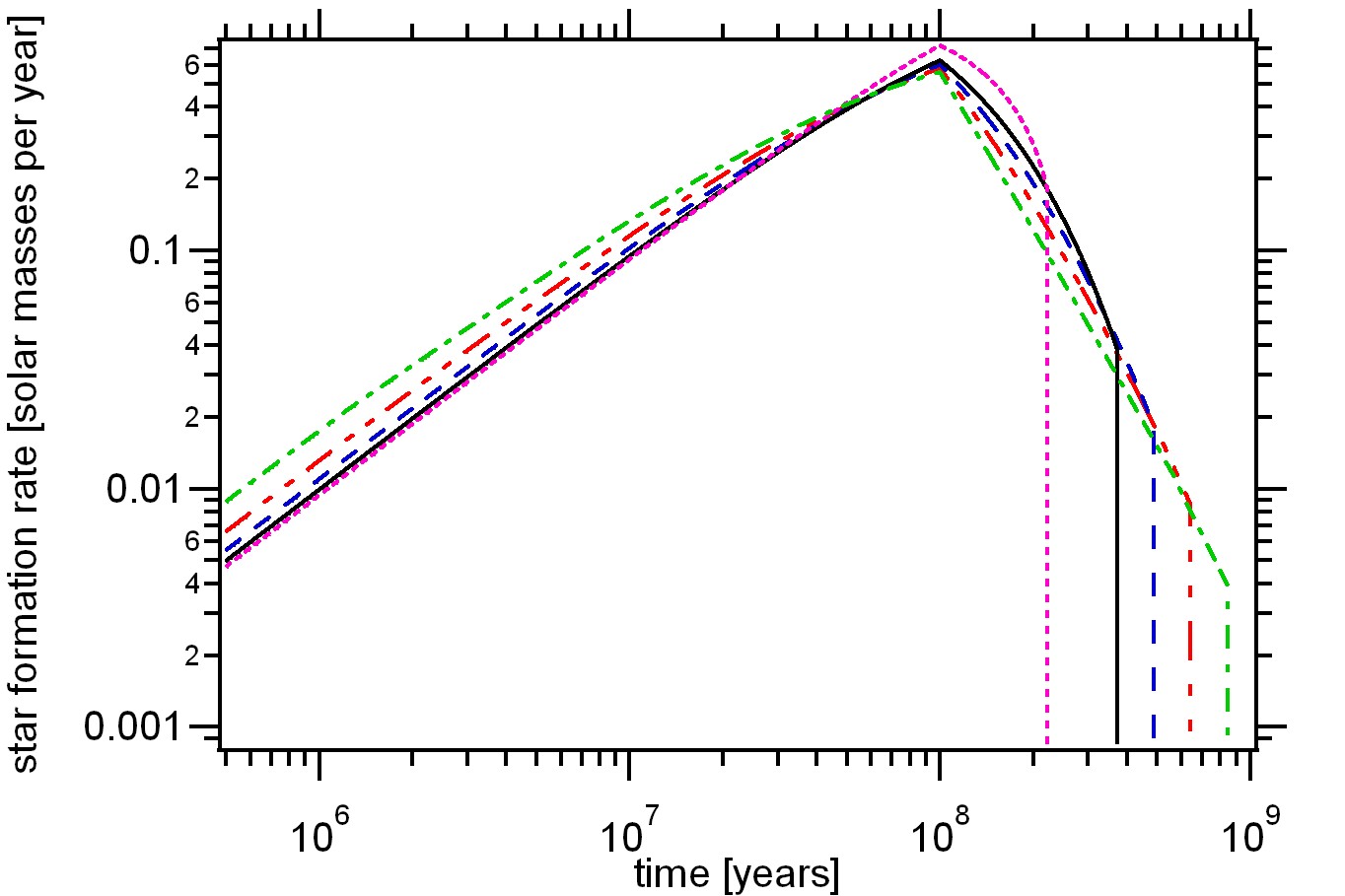}}
	\caption{Time evolution of $\dot{M}_*(t)$ for $\dot{M}_\mathrm{sup} = 1 ~M_\sun \mathrm{yr^{-1}}$ and $ t_\mathrm{sup} = 10^8 \rm{yr}$ for the different models U0 (dotted), U1 (solid), U2 (dashed), U3 (dot-dot-dashed) and U4 (dot-dashed), see further Tab.~\ref{tab:parameters_sfr}. The time at which star formation completely ceases is dependent on the model.}
	\label{fig:sfr_u_evo}
\end{figure}

\begin{figure}[bt]
	\centering
	\resizebox{\hsize}{!}{\includegraphics{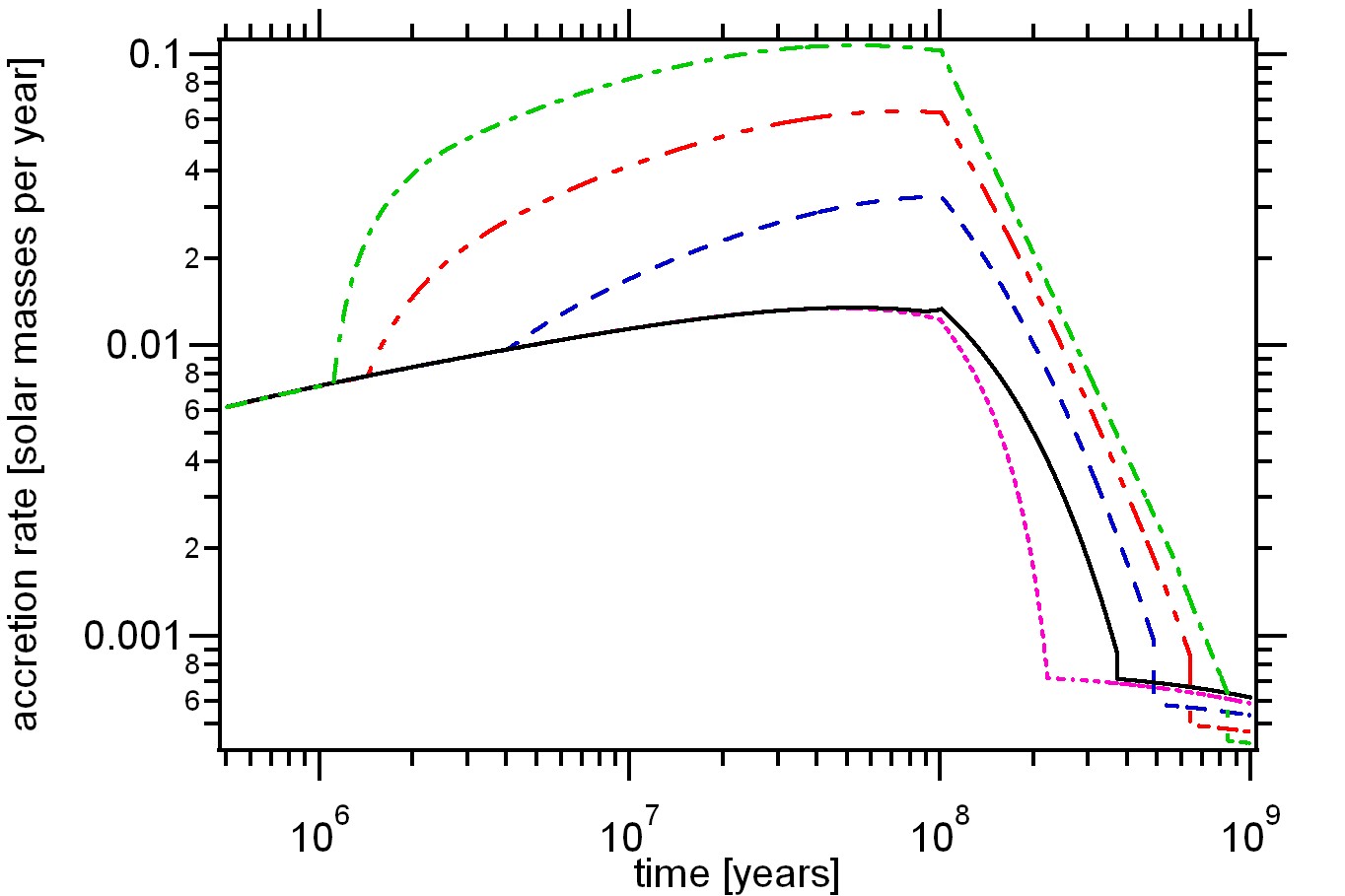}}
	\caption{Time evolution of $\dot{M}_\mathrm{BH}(t)$ for $\dot{M}_\mathrm{sup} = 1 ~M_\sun ~\mathrm{yr^{-1}}$ and $ t_\mathrm{sup} = 10^8 ~\rm{yr}$ for the different models U0 (dotted), U1 (solid), U2 (dashed), U3 (dot-dot-dashed) and U4 (dot-dashed), see further Tab.~\ref{tab:parameters_sfr}.}
	\label{fig:acc_u_evo}
\end{figure}

All models show a characteristic behaviour in the evolution of the star formation rate, in concordance with the results of \citet{Kawakatu2008}. As shown in Fig.~\ref{fig:sfr_u_evo} the overall star formation rate rises continuously as long as gas is supplied to the disk. When the supply stops, the star formation rate quickly decreases and finally drops to zero, when the circumnuclear gas disk becomes completely stable, i.e. when $r_\mathrm{c} \geq r_\mathrm{out}$. For the models with higher $\theta$ the star formation rate evolves less steeply and breaks off later than for lower $\theta$. 
\begin{figure}[tb]
	\centering
	\resizebox{\hsize}{!}{\includegraphics{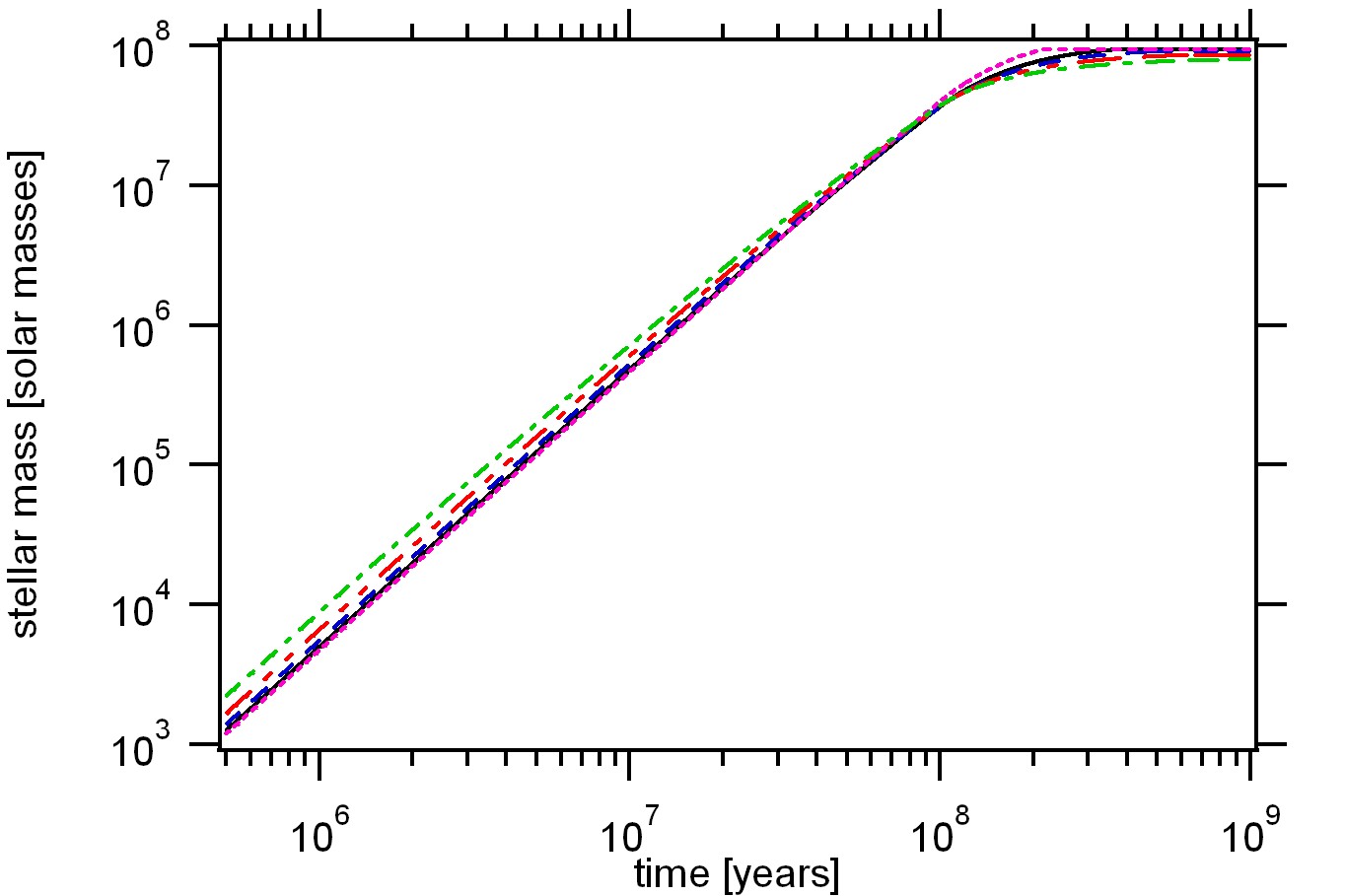}}
	\caption{Time evolution of $M_*(t)$ for $\dot{M}_\mathrm{sup} = 1 ~M_\sun ~\mathrm{yr^{-1}}$ and $ t_\mathrm{sup} = 10^8 ~\rm{yr}$ for the different models U0 (dotted), U1 (solid), U2 (dashed), U3 (dot-dot-dashed) and U4 (dot-dashed), see further Tab.~\ref{tab:parameters_sfr}. The final stellar mass is marginally dependent on the model.}
	\label{fig:mstar_u_evo}
\end{figure}
This behaviour results into similar final stellar masses for all models of approximately $ 5 \times 10^7 ~M_\sun $, as shown in Fig.~\ref{fig:mstar_u_evo}, which is a consequence of the normalization.
Although star formation rates are similar for all models the accretion rates strongly depend on the chosen star formation model. Figure~\ref{fig:acc_u_evo} shows the evolution of the accretion rates for the five studied models until time \mbox{$t = 10^9$ yr}.
For {U1} the evolution of the accretion rate resembles the results of \citet{Kawakatu2008} although the drop down to what they called \emph{low accretion phase} happens later in our model. We explain this behaviour by the fact that in our model the forming stars inject supernova-energy only $10^6$ yr after they have formed and the injected energy dissipates more slowly (see section~\ref{ssec:diskvert}), thereby delaying the ceasing of supersonic turbulence which in our model is the main driver of efficient accretion.

The obtained accretion rates of model {U1} are one order of magnitude lower than in the paper of \citet{Kawakatu2008}. {The origin of this discrepancy is not fully clear, as it remains even when we solve the equations as provided by \citet{Kawakatu2008}. The overall behavior is however similar, consisting of an efficient accretion phase during the gas supply and a significant drop when the supply is shut down.}

{We show here that the} accretion rate rises more steeply with higher $\theta$, while for very small $\theta$, the rise happens very late (for $\theta = 1$) or not at all (for $\theta = 0.5$), before the gas contents of the circumnuclear disk starts to decrease. The transition to the \emph{low accretion phase} happens significantly later than in the model by \citet{Kawakatu2008}, roughly at the same time, when star formation completely ceases. By then, almost all injected turbulent energy has been dissipated so that the drop to the low-accretion phase is not as deep.
\begin{figure}[tb]
	\centering
	\resizebox{\hsize}{!}{\includegraphics{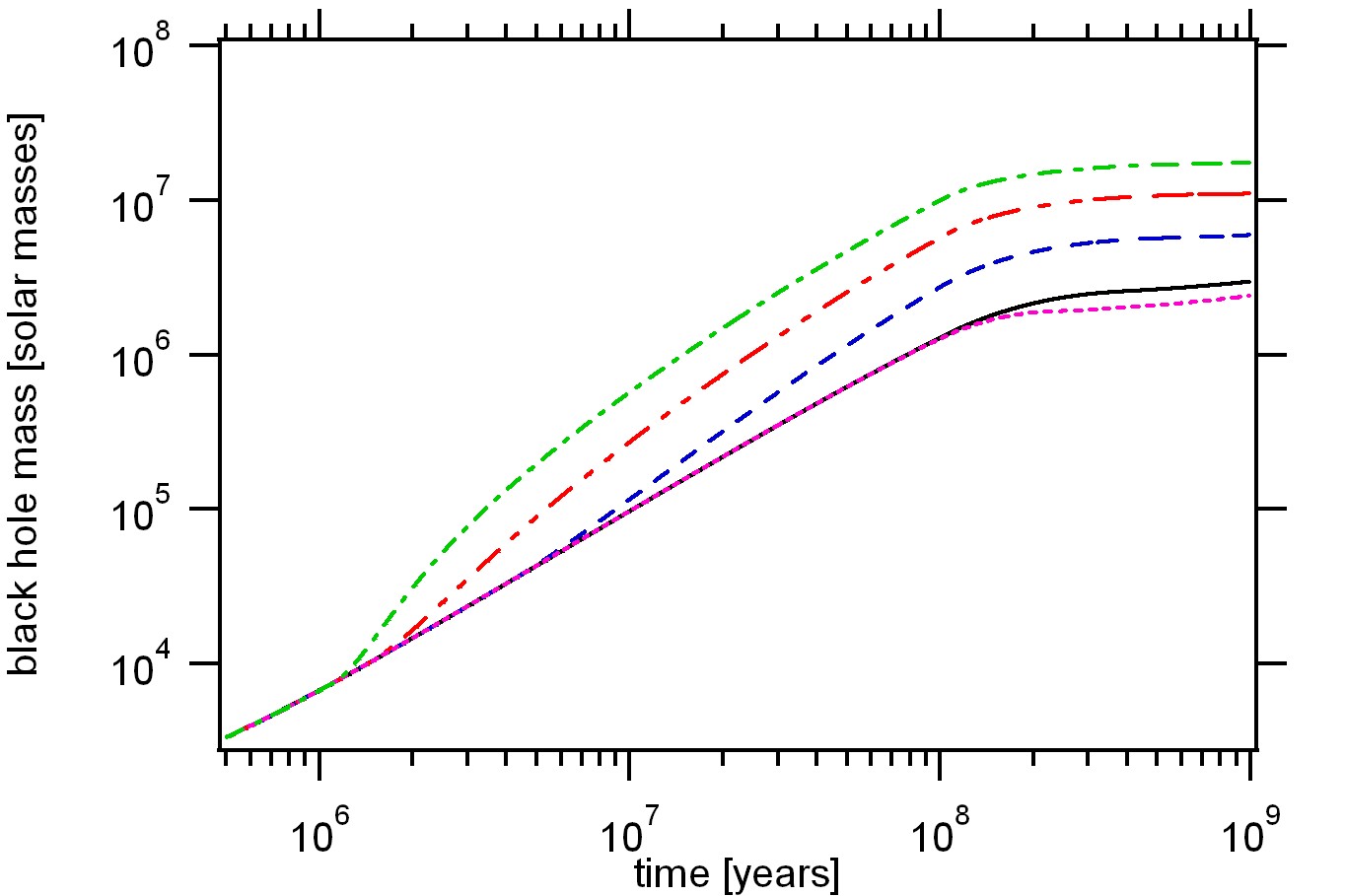}}
	\caption{Time evolution of $M_\mathrm{BH}(t)$ for $\dot{M}_\mathrm{sup} = 1 M_\sun \mathrm{yr^{-1}}$ and $ t_\mathrm{sup} = 10^8 \rm{yr}$ for the different models U0 (dotted), U1 (solid), U2 (dashed), U3 (dot-dot-dashed) and U4 (dot-dashed), see further Tab.~\ref{tab:parameters_sfr}. The final mass of the SMBH strongly depends on the model.}
	\label{fig:mbh_u_evo}
\end{figure}
The variable accretion rates result in strongly parameter-dependent final masses for the SMBH in the center of the gas disk. Different from the results of \citet{Kawakatu2008}, for small $\theta$ the black hole barely grows to $2 \times 10^6 ~M_\sun$, due to the considerably lower accretion rates. Only for larger $\theta$ the SMBH grows up to a mass of $\approx 2 \times 10^7 ~M_\sun$.\\

\begin{figure}[bt]
	\centering
	\resizebox{\hsize}{!}{\includegraphics{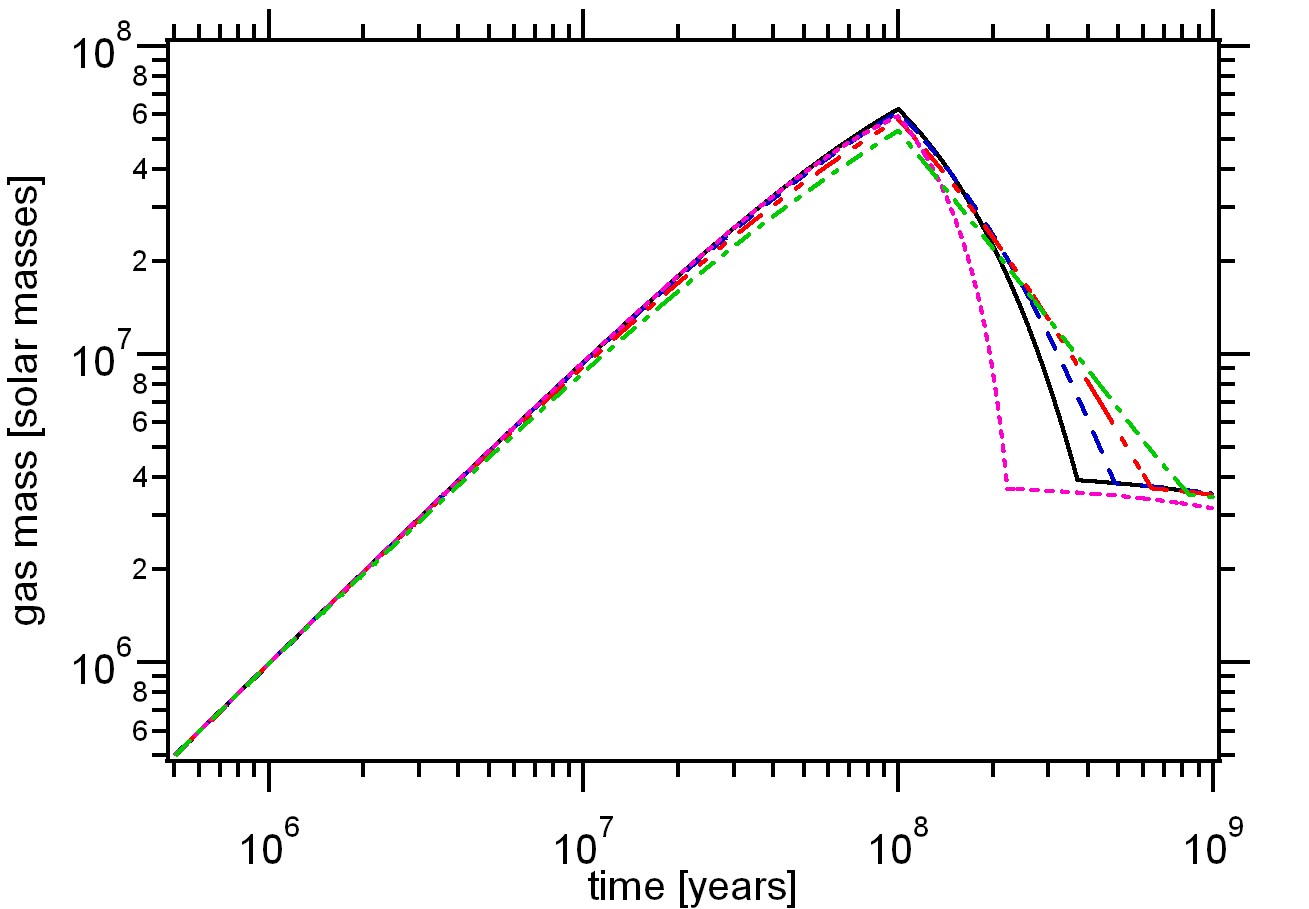}}
	\caption{Time evolution of $M_\mathrm{gas}(t)$ for $\dot{M}_\mathrm{sup} = 1 ~M_\sun ~\mathrm{yr^{-1}}$ and $ t_\mathrm{sup} = 10^8 ~\rm{yr}$ for the different models U0 (dotted), U1 (solid), U2 (dashed), U3 (dot-dot-dashed) and U4 (dot-dashed), see further Tab.~\ref{tab:parameters_sfr}. The final gas mass is marginally dependent on the model.}
	\label{fig:mgas_u_evo}
\end{figure}

The gas mass evolves similar to what \citet{Kawakatu2008} obtain, as shown in Fig.~\ref{fig:mgas_u_evo}. The gas contents of the disk are continuously increasing as long as gas is supplied from the hosting galaxy. When the supply ceases the gas mass quickly decreases, mainly due to still forming stars. As soon as star formation ceases the gas content decreases to a few $10^6 ~M_\sun$ for all models. 
\subsection{Dynamical evolution with self-regulated star formation}

For self-regulated star formation we have $\epsilon \equiv 1$ so that star formation is sensitive to an increased turbulent velocity (compare Eq.~\ref{eq:sfr_rad}). $\theta$ is a free parameter, whereas $\Psi$ has been chosen in such manner that the final stellar masses are roughly equal for all pairs of $(\theta, \Psi)$ to ensure comparability of the different models.
Tab.~\ref{tab:parameters_sfr} lists the studied models of unregulated star formation where S4 is equivalent to the model proposed by \citet{Elmegreen2010} who chose $\theta = 2$ and $\Psi = 1.4 \times 10^{-16}$.

\begin{figure}[bt]
	\centering
	\resizebox{\hsize}{!}{\includegraphics{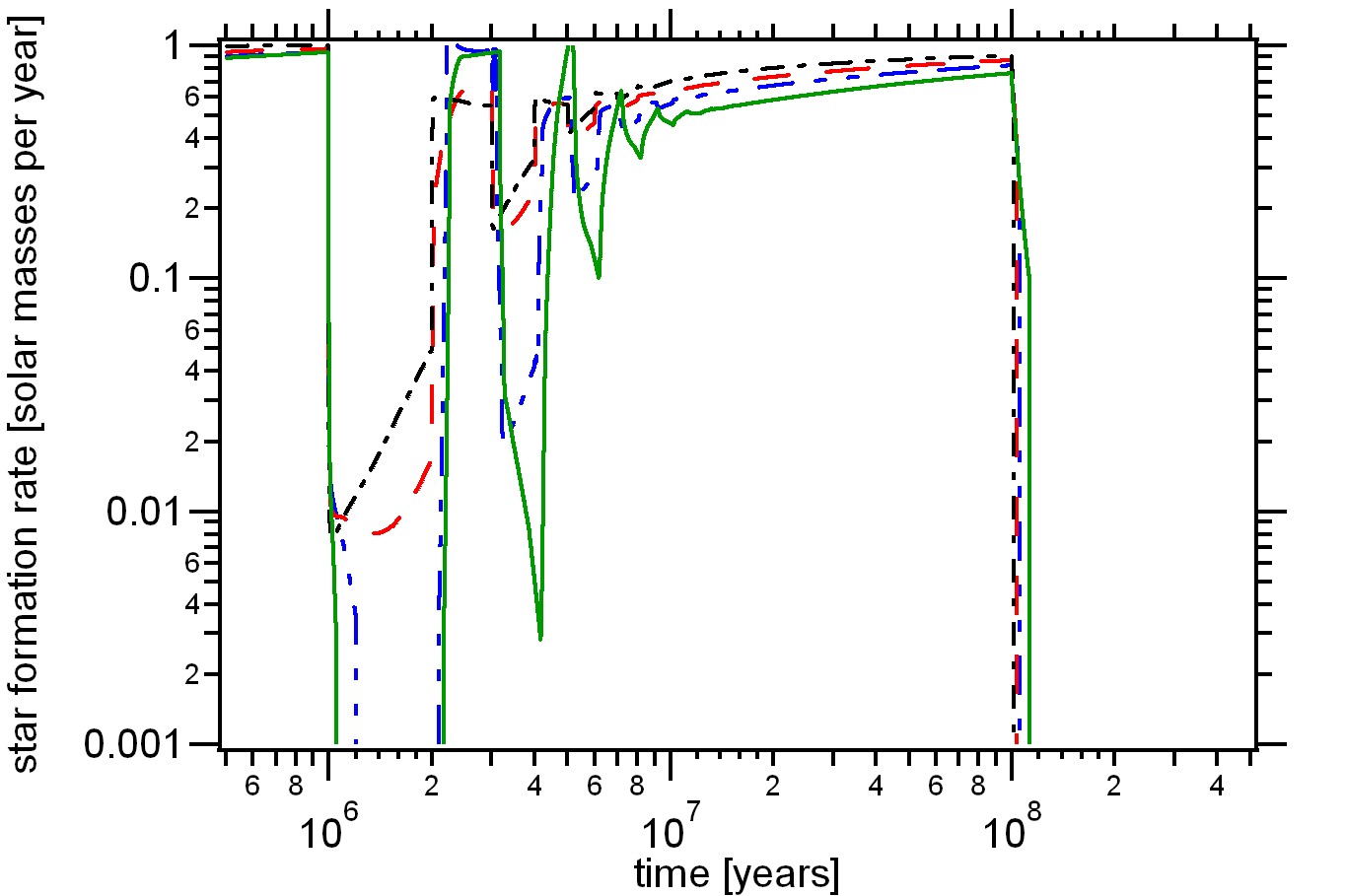}}
	\caption{Time evolution of $\dot{M}_*(t)$ for $\dot{M}_\mathrm{sup} = 1 ~M_\sun ~\mathrm{yr^{-1}}$ and $ t_\mathrm{sup} = 10^8 ~\rm{yr}$ for the different models S1 (dash-dotted), S2 (dashed), S3 (dot-dot-dashed) and S4 (solid), see further Tab.~\ref{tab:parameters_sfr}. The time at which star formation completely ceases is scarcely dependent on the model.}
	\label{fig:sfr_s_evo}
\end{figure}

The initial phase of the evolution shows some oscillatory behavior as a result of  supernova feedback, as here the injected turbulence strongly suppresses star formation activity, which reduced the feedback at later stages. However, when the evolutionary timescale of the system becomes longer than the injection timescale of the supernova feedback, these oscillations adjust and a smooth transition is obtained (see Fig.~\ref{fig:sfr_s_evo}).
 
\begin{figure}[tb]
	\centering
	\resizebox{\hsize}{!}{\includegraphics{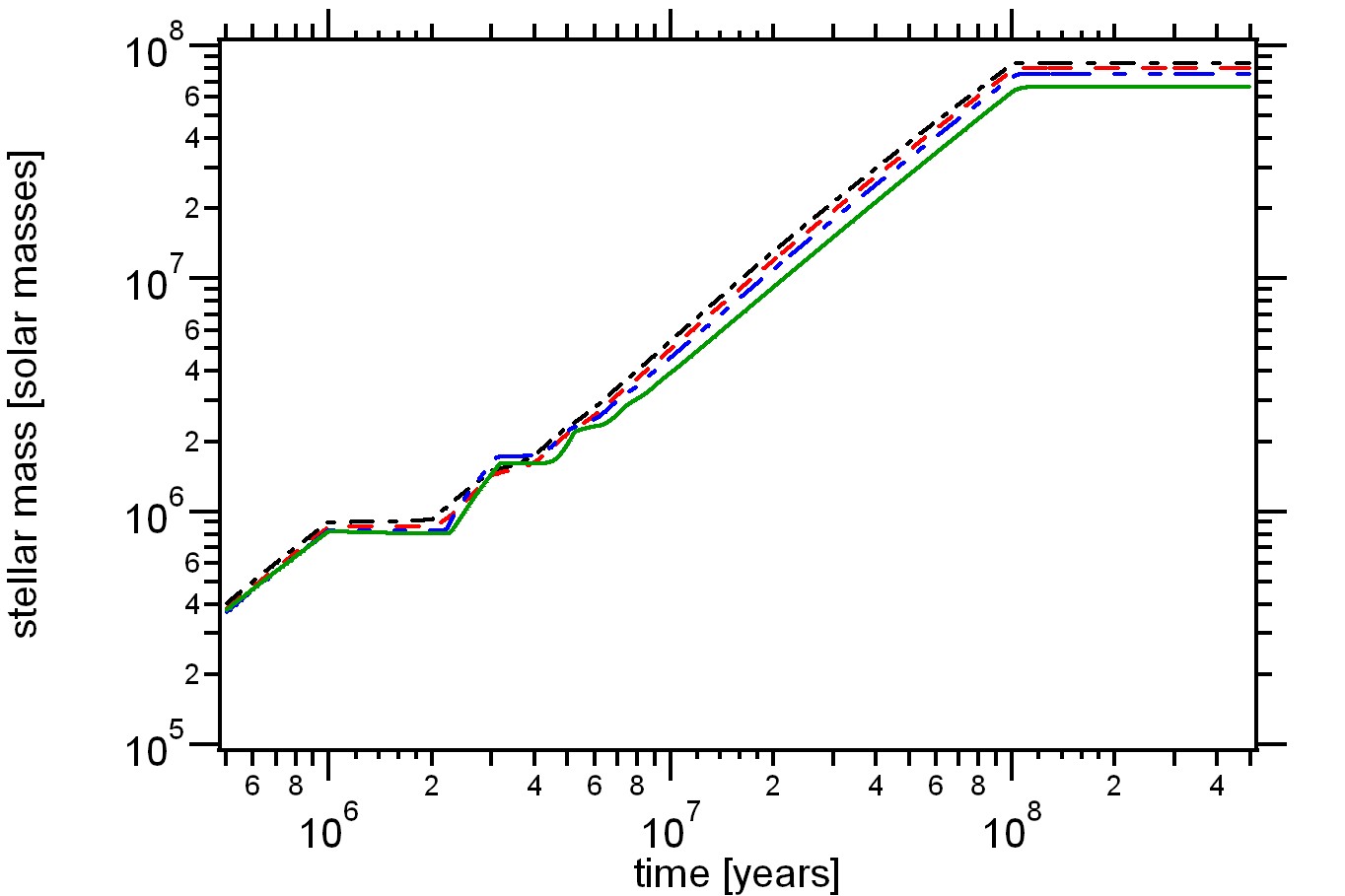}}
	\caption{Time evolution of $M_*(t)$ for $\dot{M}_\mathrm{sup} = 1 ~M_\sun ~\mathrm{yr^{-1}}$ and $ t_\mathrm{sup} = 10^8 ~\rm{yr}$ for the different models S1 (dash-dotted), S2 (dashed), S3 (dot-dot-dashed) and S4 (solid), see further Tab.~\ref{tab:parameters_sfr}.}
	\label{fig:mstar_s_evo}
\end{figure}

The star formation starts decreasing when the supply from the host galaxy stops after $10^8$~years. Even for higher $\theta$ star formation carries on only little longer ($< 10^7$ yr).
The final stellar masses are roughly $ 7 \times 10^7 ~M_\sun $ for all models as shown in Fig.~\ref{fig:mstar_s_evo}.

\begin{figure}[bt]
	\centering
	\resizebox{\hsize}{!}{\includegraphics{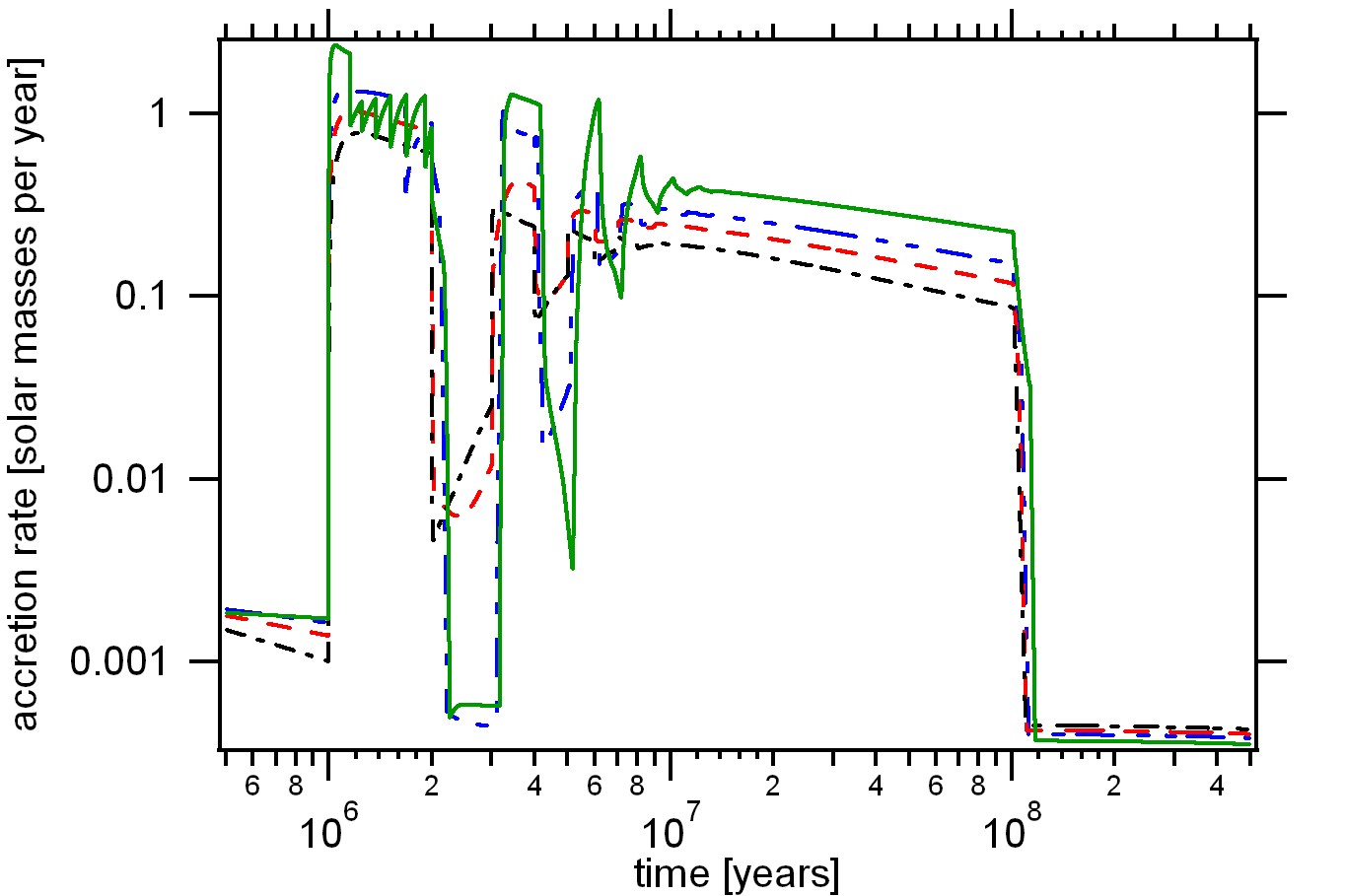}}
	\caption{Time evolution of $\dot{M}_\mathrm{BH}(t)$ for $\dot{M}_\mathrm{sup} = 1 ~M_\sun ~\mathrm{yr^{-1}}$ and $ t_\mathrm{sup} = 10^8 ~\rm{yr}$ for the different models S1 (dash-dotted), S2 (dashed), S3 (dot-dot-dashed) and S4 (solid), see further Tab.~\ref{tab:parameters_sfr}. The accretion rates are clearly dependent on model parameters.}
	\label{fig:acc_s_evo}
\end{figure}

The accretion rates clearly depend on model parameters, as can be seen in Fig. \ref{fig:acc_s_evo}. Similar to the star formation rate the accretion rates show initially an oscillatory behaviour, which is anticyclic to the star formation rate. This latter is expected, as the accretion onto the black hole is proportional to the square of the turbulent velocity (compare section~\ref{ssec:bh-acc}).

\begin{figure}[bt]
	\centering
	\resizebox{\hsize}{!}{\includegraphics{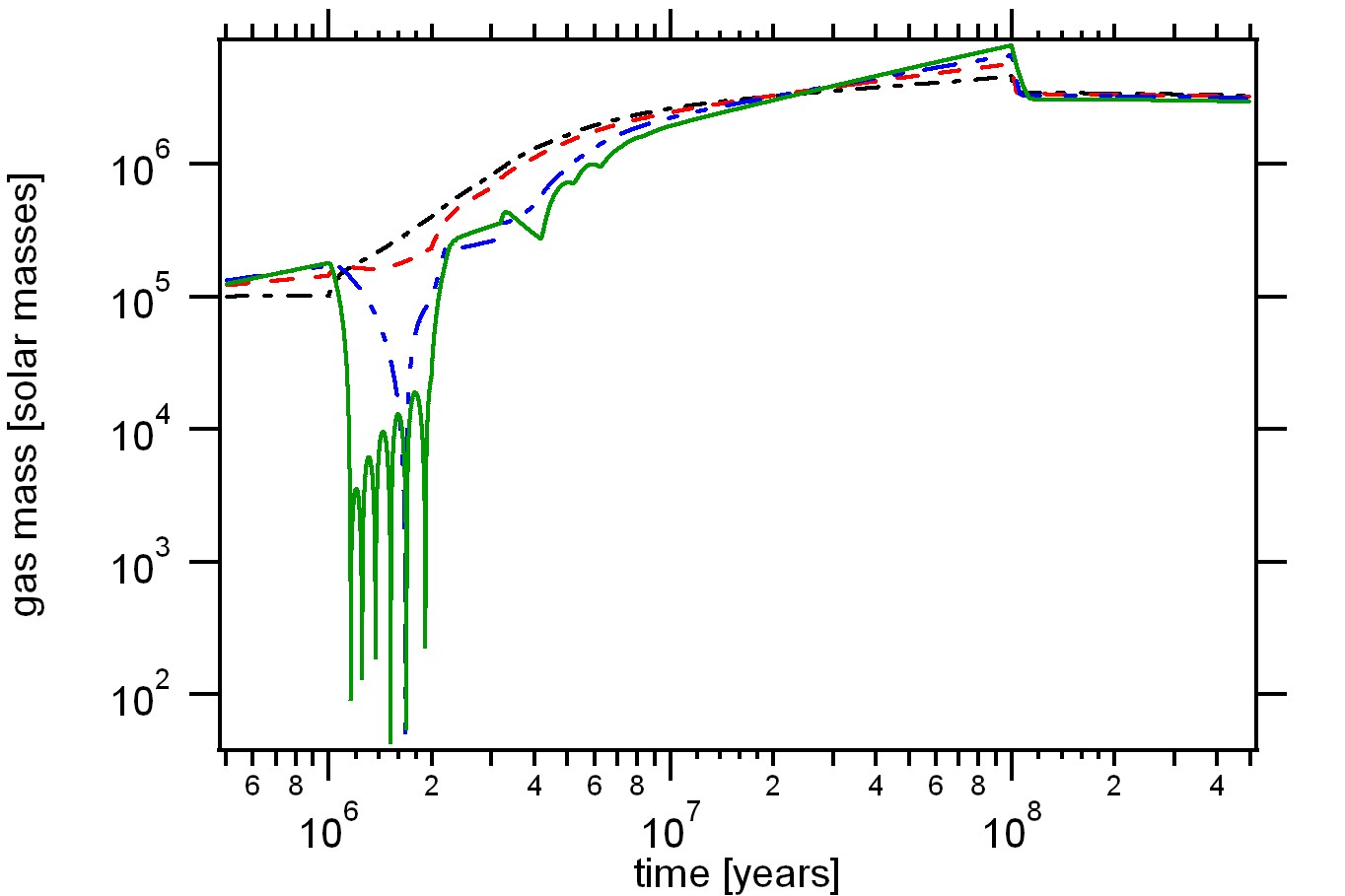}}
	\caption{Time evolution of $M_\mathrm{gas}(t) = M_\mathrm{g,i}(t) + M_\mathrm{g,o}(t)$ for \mbox{$\dot{M}_\mathrm{sup} = 1 ~M_\sun ~\mathrm{yr^{-1}}$} and \mbox{$ t_\mathrm{sup} = 10^8 ~\rm{yr}$} for the different models S1 (dash-dotted), S2 (dashed), S3 (dot-dot-dashed) and S4 (solid), see further Tab.~\ref{tab:parameters_sfr}. The final gas mass is independent of model parameters.}
	\label{fig:mgas_s_evo}
\end{figure}

Considering the accretion rates for model S4 we notice that the first oscillation is overlaid by another oscillation which is unique for this choice of parameters. The cause for the additional oscillation is the over-efficient accretion, which depletes the gas reservoir faster than it is refilled by the gas supply, which can also be seen in Fig.~\ref{fig:mgas_s_evo}. Comparing Figs.~\ref{fig:sfr_s_evo}~and~\ref{fig:acc_s_evo} it can be noticed, that the models with lower star formation during the period between $10^7$ and $10^8$~yr have the higher accretion rates during that interval. As gas masses are generally low for all models (see Fig.~\ref{fig:mgas_s_evo}), i.e. $\la$~10~\% of the supplied mass over the whole period, this seems to indicate a competition for gas between star formation and black hole accretion.

\begin{figure}[tb]
	\centering
	\resizebox{\hsize}{!}{\includegraphics{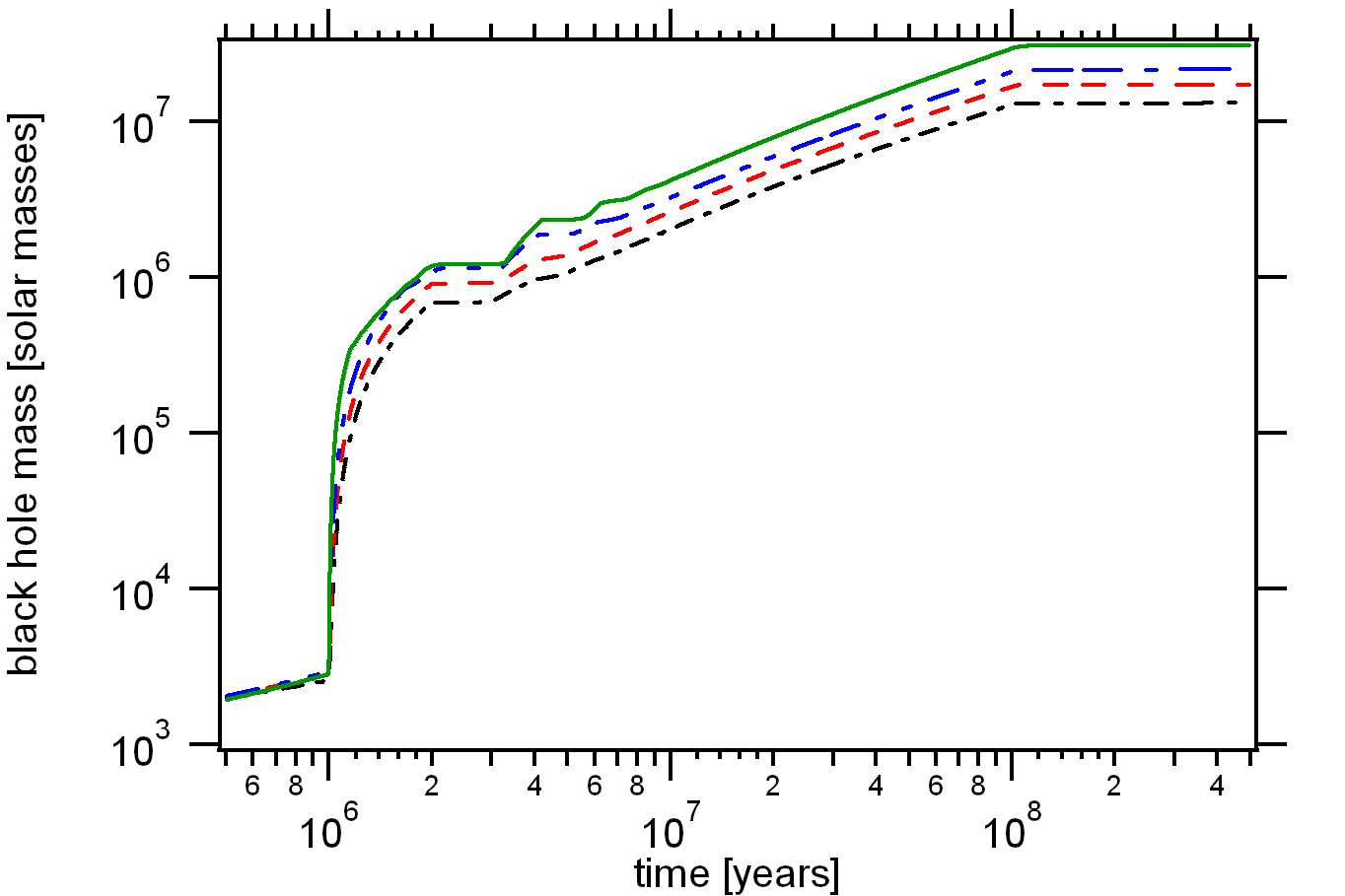}}
	\caption{Time evolution of $M_\mathrm{BH}(t)$ for $\dot{M}_\mathrm{sup} = 1 ~M_\sun ~\mathrm{yr^{-1}}$ and $ t_\mathrm{sup} = 10^8 ~\rm{yr}$ for the different models S1 (dash-dotted), S2 (dashed), S3 (dot-dot-dashed) and S4 (solid), see further Tab.~\ref{tab:parameters_sfr}. The final black hole mass is clearly dependent on model parameters.}
	\label{fig:mbh_s_evo}
\end{figure}

For the different models we obtain final black hole masses of a few $10^7 ~M_\sun$ with higher masses for greater $\theta$. As shown in Fig.~\ref{fig:mbh_s_evo} the dependence of the black hole mass on model parameters is significant but not as strong as for unregulated star formation models (compare Fig.~\ref{fig:mbh_u_evo}).

The evolution of the gas masses is slightly dependend on model parameters as shown in Fig.~\ref{fig:mgas_s_evo}. However, the final gas masses are insensitive to the choice of parameters and amount to approximately $5 \times 10^6 ~M_\sun$ for all models.

For model S4 we observe several oscillations between \mbox{$ t = 10^6 $ yr} and \mbox{$ t = 2 \times 10^6 $ yr} which are caused by the over-efficient accretion onto the black hole as mentioned above. In the same period the gas mass of model S3 shows a single dip, supposedly for the same reasons. For low $\theta$ we do not observe such an \emph{overreaction} to the onset of supernova-feedback, because the star formation per unit area follows a less steep power law which results into a less steep power law for the turbulent velocity and therefore also for the scale height. As the accretion rate is linear dependend on the turbulent velocity as well as the scale height at the inner disk radius $r_\mathrm{in}$, where both reach their highest values, the accretion rate reacts strongly to a steeper power law for velocity and scale height.

Table \ref{tab:s-u-masses_comp} gives an overview of the mass balance for all models. As expected black hole masses rise with higher $\theta$ as steeper power laws for the gas surface density and thereby for the star formation per unit area result into higher accretion rates onto the central black hole. The disk mass is reduced accordingly where we like to mention that most of the final disk mass is contributed by stars, as final gas masses are smaller than stellar masses by approximately one order of magnitude for all models (compare Figs. \ref{fig:mstar_u_evo} with \ref{fig:mgas_u_evo} and \ref{fig:mstar_s_evo} with \ref{fig:mgas_s_evo}).
\begin{table}[tb]
	\centering
	\caption{Final masses for the black hole and the disk for all models}
	\begin{tabular}{cccc}\hline\hline\addlinespace[0.1cm]
	model	&	$M_\mathrm{BH}$			&	$M_\mathrm{disk}$		& $ M_\mathrm{BH} / M_\mathrm{disk} $\\
			&	$(10^7 M_\sun)$	&	$ (10^7 M_\sun) $	&	 \\\hline\addlinespace[0.1cm]
	U0		&		0.24	&		9.76	&	0.02	\\
	U1		&		0.30	&		9.70	&	0.03	\\
	U2		&		0.59	&		9.41	&	0.06	\\
	U3		&		1.10	&		8.90	&	0.12	\\
	U4		&		1.74	&		8.26	&	0.21	\\\hline\addlinespace[0.1cm]
	S1		&		1.32	&		8.68	&	0.15	\\
	S2		&		1.73	&		8.37	&	0.21	\\
	S3		&		2.16	&		7.84	&	0.28	\\
	S4		&		3.00	&		7.00	&	0.43	\\\hline
	\end{tabular}
	\tablefoot{The first column identifies the model (see Tab. \ref{tab:parameters_sfr}), second and third column denote the mass of the black hole and the disk. The third column displays the mass ratio $M_\mathrm{BH} / M_\mathrm{disk}$.}
	\label{tab:s-u-masses_comp}
\end{table}

\subsection{Radial disk structure}
\begin{figure}[bt]
	\centering
	\resizebox{\hsize}{!}{\includegraphics{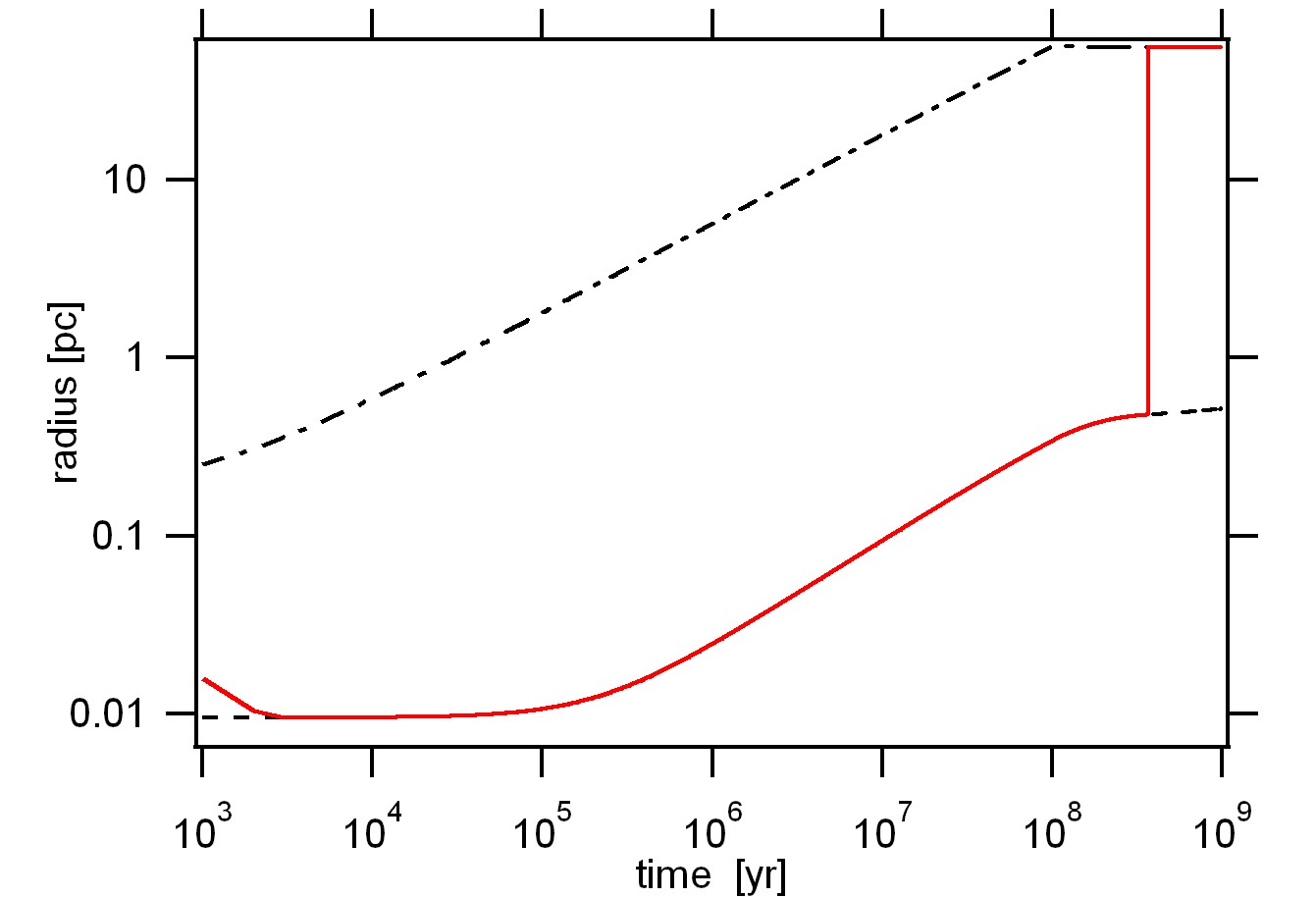}}
	\caption{Evolution of radial structure with the outer radius $r_\mathrm{out}$ (dash-dotted), the inner radius $r_\mathrm{in}$ (dashed) and the critical radius $r_\mathrm{c}$ (solid) for model U1.}
	\label{fig:radii_u_evo}
\end{figure}

\begin{figure}[tb]
	\centering
	\resizebox{\hsize}{!}{\includegraphics{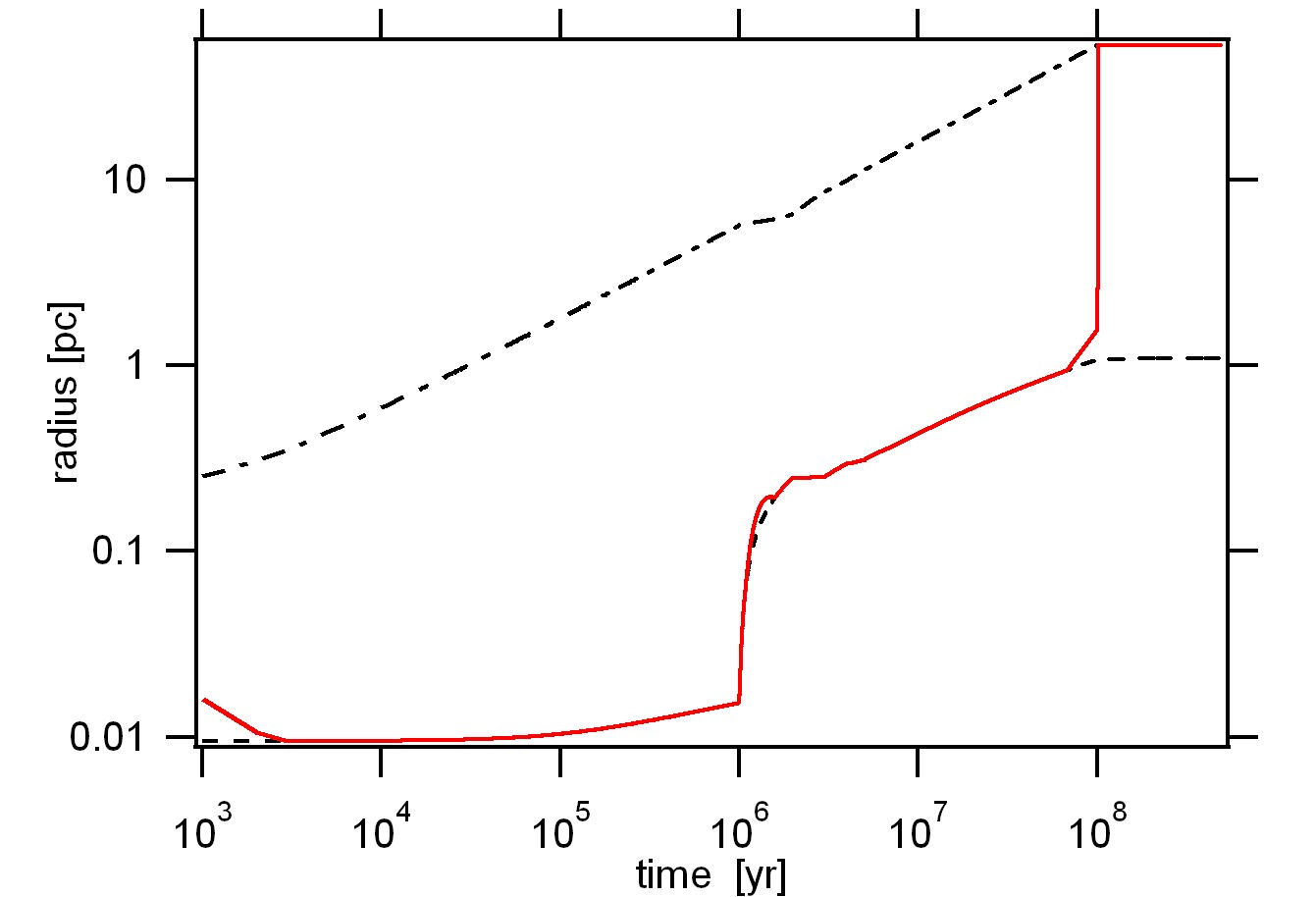}}
	\caption{Evolution of radial structure with the outer radius $r_\mathrm{out}$ (dash-dotted), the inner radius $r_\mathrm{in}$ (dashed) and the critical radius $r_\mathrm{c}$ (solid) for model S1.}
	\label{fig:radii_s_evo}
\end{figure}

The evolution of the characteristic radii is quite similar for the different star formation models. While the outer radius for all models is dominated by the disk mass, therefore by the matter supply from the hosting galaxy, the inner radius more or less traces the growth of the SMBH (see sec. \ref{ssec:diskrad}). As can be seen from Figs. \ref{fig:radii_u_evo} and \ref{fig:radii_s_evo} most of the time the circumnuclear disk is either completely self gravitating or completely stable against self-gravitation. The self-regulating star formation seems to result into longer transitional phases where the critical radius $r_\mathrm{c}$ lies inside the disk, especially after the gas supply from the host galaxy has ceased.
\section{Discussion of results and comparison to observations}
\subsection{Influence of the mass supply on the black hole growth} 
We studied the influence of the mass supply from the host galaxy on the final black hole mass and stellar mass of the system. We applied supply rates $\dot{M}_\mathrm{sup} = 1 \dots 100 \ M_\sun~\mathrm{yr^{-1}}$ exemplary on the four models U1, U4, S1 and S4 (see Table \ref{tab:parameters_sfr}) with a run-time of $10^9$ yr.
\begin{figure}[tb]
	\centering
	\resizebox{\hsize}{!}{\includegraphics{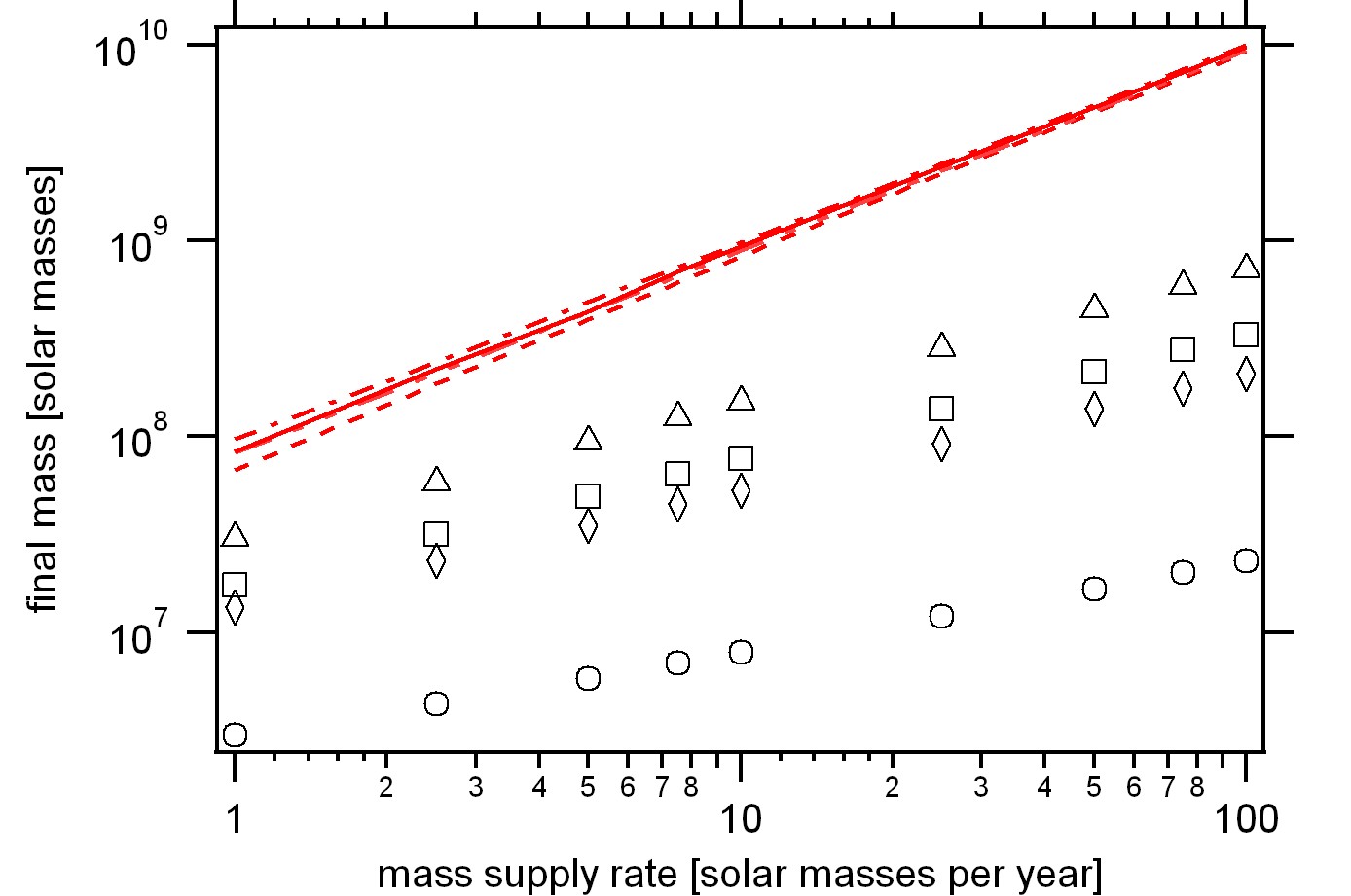}}
	\caption{Final masses of the black hole (markers) and the stars (lines) in dependence of the mass supply for the following models: U1 (circle and dot-dashed), U4 (square and dot-dot-dashed), S1 (diamond and solid), S4 (triangle and dashed). For better comparison all models were evolved until $t = 10^9$ yr.}
	\label{fig:pstudy_psup}
\end{figure}
As displayed in Fig.~\ref{fig:pstudy_psup} final stellar masses as well as final black hole masses follow a power law \mbox{$M_\mathrm{final} = M_0 \cdot (\dot{M}_\mathrm{sup}/[M_\sun\mathrm{yr^{-1}}])^x$} with model-dependent exponent $x$. The stellar mass $x$ is roughly the same for all models whereas for the black hole masses the exponents differ significantly for the various models, as shown in Table~\ref{tab:pstudy_psup}.

\begin{table}[tbh]
	\caption{Dependence of final stellar and black hole masses on mass supply rate $\dot{M}_\mathrm{sup}$}
	\begin{tabular}{cccc}\hline\hline\addlinespace[0.1cm]
	model	&	$x(M_\mathrm{BH})$	&	$x(M_*)$	& $x(M_\mathrm{BH}/M_*)$\\\hline\addlinespace[0.1cm]
	U1		&	$0.48$				&	$1.00$		& $-0.60$\\
	U4		&	$0.62$				&	$1.01$		& $-0.37$\\
	S1		&	$0.59$				&	$1.01$		& $-0.45$\\
	S4		&	$0.66$				&	$1.03$		& $-0.42$\\\hline
	\end{tabular}
	\label{tab:pstudy_psup}
	\tablefoot{first column: studied model, for model parameters see Table \ref{tab:parameters_sfr}; second and third column: power law-exponents $x$ for the final black hole mass and the final stellar mass, see also Fig.~\ref{fig:pstudy_psup}; fourth column: power law-exponents for the mass ratio $M_\mathrm{BH} / M_*$}
\end{table}
In general, the final stellar masses are approximately linearly dependend on the mass supply rate for all models. The dependence of the final black hole masses on the mass supply rate is weaker and differs significantly between self-regulated star formation and unregulated star formation (compare especially models U1 and S1 with $\theta = 1.0$ for both models) as well as between different $\theta$. We note that the exponent is rising with rising $\theta$, but more steeply for the unregulated star formation (see Table~\ref{tab:pstudy_psup}). Consequently the mass ratio $M_\mathrm{BH} / M_*$ depends on the mass supply rate as well.
\subsection{Influence of the seed mass}
As there is a great discussion in the literature about the possible seeds of SMBHs and processes to form very massive seeds, e.g. \citet{Davies2011, Johnson2013, Latif13a, Latif13b, Schleicher13, Borm13}, we studied the influence of the seed mass on the final black hole mass.
For seed masses from $100~M_\sun$ to $10^6~M_\sun$ we evolved the models U1 and S1 (see Table \ref{tab:parameters_sfr}) for 1~Gyr and compared final masses of the SMBH and the stars. The results as shown in Table \ref{tab:pstudy_mseed} make clear, that in our model even very massive seeds do not enhance the growth of the SMBH. This is in agreement with the findings of \citet{Montesinos2011}, who found that the seed mass influenced the final SMBH mass only weakly, albeit they did not include star formation into their simulation.

\begin{table}[tbh]
	\caption{Dependence of final stellar and black hole masses on the mass of the black hole seed}
	\begin{tabular}{ccccc}\hline\hline\addlinespace[0.1cm]
	 			&	\multicolumn{2}{c}{model U1}	&	\multicolumn{2}{c}{model S1}	\\
	seed mass	& $M_\mathrm{BH}$	&	$M_*$			& $M_\mathrm{BH}$	&	$M_*$	\\
	$(M_\sun)$	& $(10^6 M_\sun)$	& $(10^7 M_\sun)$	& $(10^7 M_\sun)$	& $(10^7 M_\sun)$	\\ \hline \addlinespace[0.1cm]
	$10^2$		&	$2.96$			&	$9.36$			& $1.34$			&	$8.36$	\\
	$10^3$		&	$2.96$			&	$9.36$			& $1.34$			&	$8.36$	\\
	$10^4$		&	$2.96$			&	$9.36$			& $1.34$			&	$8.36$	\\
	$10^5$		&	$3.01$			&	$9.36$			& $1.34$			&	$8.36$	\\
	$10^6$		&	$3.74$			&	$9.38$			& $1.38$			&	$8.40$	\\\hline
	\end{tabular}
	\label{tab:pstudy_mseed}
	\tablefoot{first column: mass of black hole seed; second and third column: final masses of the black hole and the stars for model U1; fourth and fifth column: final masses of the black hole and the stars for model S1; for model parameters see Table \ref{tab:parameters_sfr}}
\end{table}

In contrast to expectations very massive seeds even result into a slightly reduced amount of accreted matter and at the same time slightly enhance star formation. The reason for this non-intuitive behaviour lies in the considerably larger disk, as a very massive seed black hole dominates the gravitational potential on larger scales. This means, that the gaseous disk from the very beginning reaches out to several parsecs rather than to a few \mbox{$10^{-1}$ pc}, which results into considerably lower gas surface densities at the inner radius, as the same amount of gas is distributed over a larger disk area. As the accretion rate is linearly dependend on the gas surface density the larger disk leads to accretion rates which are smaller by two orders of magnitude. These differences hold until the gravitational potential is dominated by the disk mass, i.e. until $M_\mathrm{disk} > M_\mathrm{BH}$.

However, we note here that the environmental conditions of the first seed black holes are still unclear, and  the approximations employed in this study are not necessarily appropriate for the growth of low-mass black holes in highly metal-poor environments. The growth of seed black holes provided by different mechanisms should ideally be pursuit in numerical simulations which account for the specific conditions in the local environments.
\subsection{Nuclear star formation law: implications from observations}

{As our calculations have shown that the black hole growth depends significantly on the adopted star formation law, we explore whether the latter can be constrained from the detailed observations available for NGC~1097. We further discuss the observational status of other systems to assess whether significant restrictions for our model can be obtained.}

\subsubsection*{NGC~1097}
%
%
In order to evaluate the ability of our parametrized star formation {law} to calculate realistic star formation rates, we applied it to observational data of NGC~1097 \citep{Hsieh2011}, a nearby Seyfert 1 galaxy with a prominent star burst ring, and compared the obtained results with the observed star formation rates.

The authors gave estimates of the gas masses of individual \emph{giant molecular cloud associations} (GMAs), their central coordinates, diameters and velocity distributions as well as star formation rates per unit area, deduced from the Pa$\alpha$-luminosities  \citep{Hsieh2011}.
To apply the star formation model we considered annuli covering the complete region up to a radius of 1500~pc, and calculated the contribution of each GMA to the different annuli as well as the average velocity distribution $v_\mathrm{\bf o}$. We then calculated a theoretical star formation {rate} per unit area for each annulus with inner radius $R_{in}$ and outer radius~$R_\mathrm{out}$:

\begin{equation}
 \dot{\xi}_* = \Psi \, \Sigma_\mathrm{\bf o}^\theta \cdot  v_\mathrm{\bf o}^{-\epsilon}
\label{eq:NGC1097_modelsfr}
\end{equation}

\begin{figure}[tb]
 \centering
 \resizebox{\hsize}{!}{\includegraphics{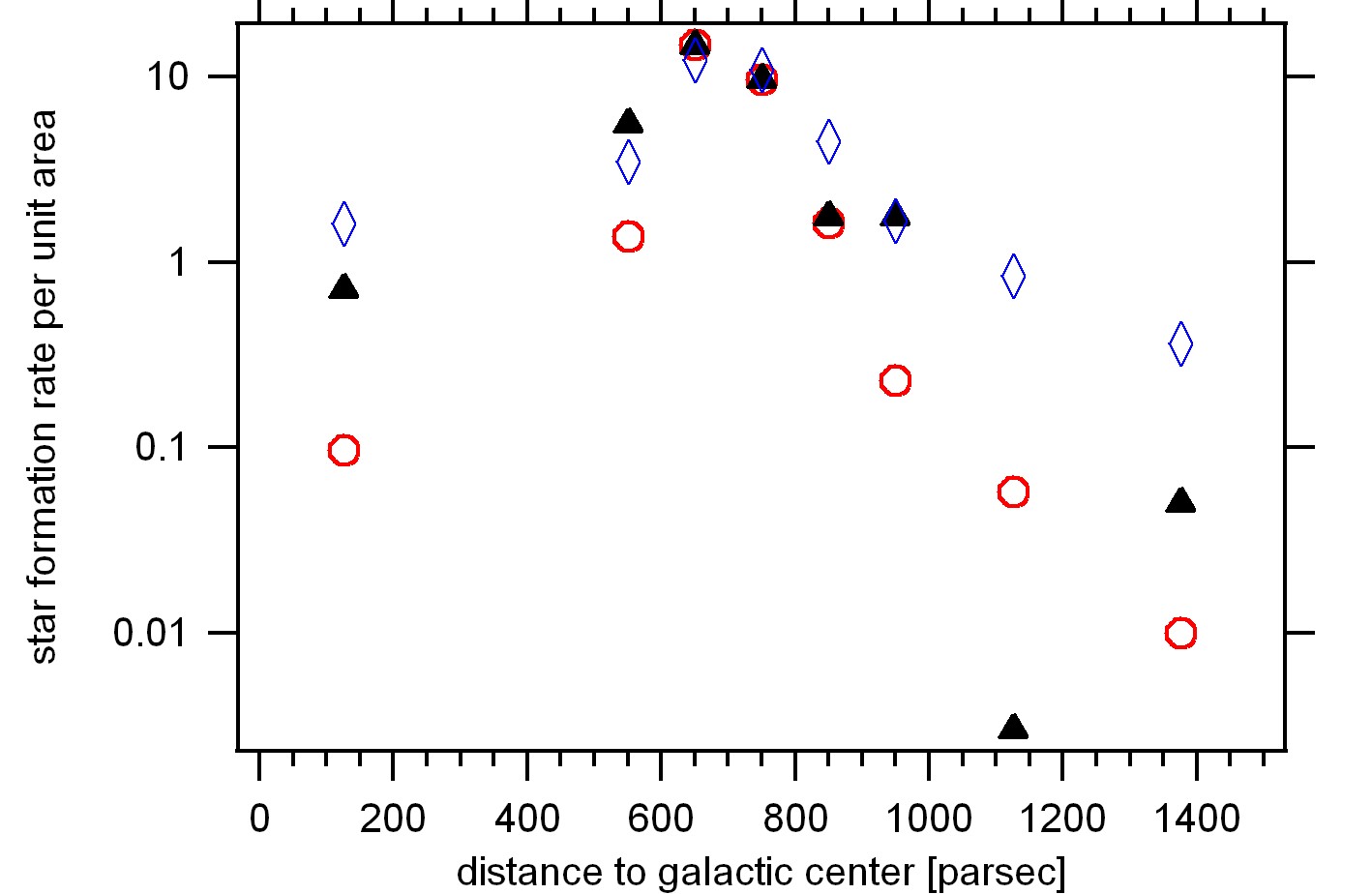}}
 \caption{Star formation rate per unit area in units of $M_\sun \, \mathrm{yr^{-1} \, kpc^{-2}}$ in NGC~1097: observed values (triangle) from \citet{Hsieh2011} and modeled values for the models U1 (diamond) and S2 with $\Psi = 7 \cdot 10^{-16}$ (circle), see Table \ref{tab:parameters_sfr}, corresponding to the models of \citet{Kawakatu2008} and \citet{Elmegreen2010} with adjusted star formation efficiencies (see text). For the innermost and the outer radii the annuli have a width of 250 pc, in the region of the star burst ring the width is only 100 pc.}
 \label{fig:NGC1097_sfr}
\end{figure}

{Here,} $\Sigma_\mathrm{\bf o}$ denotes the gas surface density in the annulus, calculated from the contributions of the GMAs to the different annuli divided by the area of the annulus $A_\mathrm{\bf o}=\pi (R_\mathrm{out}^2-R_\mathrm{in}^2)$.
In \mbox{Fig. \ref{fig:NGC1097_sfr}} we compare the observational data with the two different star formation models U1 and S4 with $\Psi = 7 \times 10^{-16}$ (see Table \ref{tab:parameters_sfr}), corresponding to the models of \citet{Kawakatu2008} with a third of their star formation efficiency (diamond) and \citet{Elmegreen2010} with a star formation efficiency of 5~\% (circle) instead of 1~\%.

Both star formation models roughly trace the observed values for the anuli with high star formation rates. Annuli with very low rates cannot be reproduced well by both models. This may indicate that the star formation efficiencies vary for the different annuli.
It is noticable that, in order to fit the models to the observed values, we have to apply a smaller star formation efficiency for the model of \citet{Kawakatu2008}, where the model of \citet{Elmegreen2010} needs to be supplied with a larger star formation efficiency than originally proposed. The latter is not surprising, as we consider a starburst region where the star formation efficiency should be higher than average, {while} the star formation model of \citet{Elmegreen2010} was suggested for general star formation in galactic disks.

For NGC~1097 \citet{Davies2007} estimated the dynamical mass in the central 20 - 40 pc from the kinematics of the stars and gas to $ 1.4 \times 10^8 ~M_\sun $, where $ 1.2 \times 10^8 ~M_\sun $ are contributed by the central black hole \citep{Lewis2006}. 
This implies a mass of gas and stars of $ 2 \times 10^7 ~M_\sun $. This results into a mass ratio of $ M_\mathrm{BH} / M_* \approx 25$. On scales of $250$~pc, \citet{Hsieh2011} give a similar value of $3.2\times10^8 \times ~M_\sun$ for the molecular gas, while the stellar mass corresponds to $\sim3.0\times10^9$~M$_\sun$. {The ratio of stellar mass to disk mass is then of the order $\sim30$, which is consistent with the expectations in most of our models except U1 (see Fig.~\ref{fig:pstudy_psup}).}

\subsubsection*{Other objects}
\begin{table*}[tb] 
	\centering
	\caption{Masses and accretion rates in nuclei of chosen AGN}  
	\begin{tabular}{c c c c c c c}\hline\hline\addlinespace[0.1cm]
object		&	$\log(M_\mathrm{BH}/M_\sun)$	&	$\log(M_\mathrm{disk}/M_\sun)$	&	radius/pc& $M_\mathrm{BH} / M_\mathrm{disk}$ 	& $L_\mathrm{acc} / L_\mathrm{Edd}	$	&	$\dot{M}_\mathrm{BH} / M_\sun$\\\hline\addlinespace[0.1cm]
	NGC~1068	&	6.94 (1)	&	8.08 (2)					&	35 (2)	&	0.07		&$0.339$ (1) & $3.06\times10^{-2}$\\
	NGC~3227	&	7.59 (1)	&	7.30\tablefootmark{a} (3)	&	63 (3)	&	1.95		&$0.022$ (1) & $8.88\times10^{-3}$\\
	NGC~3783	&	7.47 (1)	&	7.30 - 7.85  (2)			&	60 (2)	&	1.48 - 0.42	&$0.053$ (1) & $1.62\times10^{-2}$\\
	NGC~7469	&	7.09 (1)	&	7.56\tablefootmark{a} (4)	&	65 (4)	&	0.34		&$0.157$ (1) & $2.00\times10^{-2}$\\ \hline
	\end{tabular}
	\tablefoot{
	For each object (column 1), in column 2 the black hole mass is given as estimated from reverberation mapping except for NGC 1068, for which the dynamical mass estimate is given. 
	Column 3 gives the disk mass inside the radius denoted in column 4, either estimated from the dynamical mass or from the K-band luminosity. 
	Column 5 gives the efficiency of accretion luminosity $\epsilon_\mathrm{L} = L_\mathrm{acc} / L_\mathrm{Edd}$ from observation, from which the accretion rate $\dot{M}_\mathrm{BH} = \epsilon_\mathrm{L} / \epsilon_\mathrm{M} \cdot (1-\epsilon_\mathrm{M}) \cdot M_\mathrm{BH} / \tau$ can be calculated, where $\epsilon_\mathrm{M} $ is the mass accretion efficiency with a typical value of $0.3$ and $\tau$ is the characteristical accretion timescale with a value of $\approx 0.23~\mathrm{Gyr}$  \citep[see][]{Shapiro2005}.
	\tablefoottext{a}{stellar mass only}
	}
	\tablebib{
	(1) \citet{Khorunzhev2012}; 
	(2) \citet{Davies2007}; 
	(3) \citet{Davies2006};
	(4) \citet{Davies2004} 
	}
	\label{tab:obs_CND}
\end{table*}
For different AGN, a certain range of black hole - to stellar mass ratios has been reported in the literature, as detailed in Table \ref{tab:obs_CND}. NGC~1068 is a case of a relatively small SMBH in the center of a very massive disk with a mass ratio similar to those we obtain for our models with unregulated star formation.
\citet{Davies2007} found that this AGN shows no current star formation but had experienced a star burst $\approx 2-3 \times 10^8$ yr ago {on} a possible time scale of $10^8$ yr. This starburst must have bound the majority of the available gas thereby producing the very small black hole-to-stellar mass ratio, which is similar to those we obtained for our model with unregulated star formation (i.e. $\epsilon = 0$).

For NGC~3227 the comparison with observational data is more difficult, as the gas mass estimate is connected with great uncertainties of $2-20 \times 10^7 M_\sun$ where the stellar mass itself is estimated to {be} $2 \times 10^7 M_\sun$ \citep{Davies2006}. In a later paper the authors suggest that the starburst was confined to the inner 12 pc around the SMBH \citep{Davies2007}. This means, that the gas component might not belong to the original circumnuclear disk. Therefore we compare the black hole mass only to the stellar component of the dynamical mass estimate. The resulting mass ratio is considerably higher than those obtained by our models. Still, the black hole mass of NGC~3227 as well as the stellar mass are both of the same order as the final masses of model S4 (see Table~\ref{tab:s-u-masses_comp}).

For NGC~3783 the estimate of the disk mass is better constrained but still allowing for several interpretations \citep{Davies2007}. Assuming the higher value of $ 7\times 10^7 M_\sun$ for the disk mass gives a mass ratio consistent with our results for model S4. For the smallest disk mass estimate the mass ratio is comparable to that for NGC~3227, which would be inconsistent with our model. An observational verification to distinguish between these scenarios would thus be valuable for our understanding of black hole accretion.

For NGC~7469 the stellar component was estimated from the stellar $K$-band luminosity \citep{Davies2007}.
The black hole-to-stellar mass ratio is comparable to those obtained by our models with self-regulated star formation as are the absolute masses. \citet{Davies2007} suggest an age of $10^8$ yr for the most recent star burst with a duration of approximately $10^7$ yr where the authors argue against continuous star formation. We do not see such behaviour in our models, which assume a constant mass supply rates during the first $10^8$~years. Their observations might thus indicate a strong variation of the mass supply in this particular galaxies, which could produce such an intense and short-lived starburst. How such a large stellar mass could build up during the available time may indeed be a subject of further investigation. 

For the observed local AGN the accretion rates are of similar order of $10^{-2} ~M_\sun ~\mathrm{yr^{-1}}$, where in our models accretion rates can be of order unity when accretion is efficient. As none of the discussed objects are currently in a starburst phase, such a discrepancy may be expected. In fact, our models for the post-starburst phase show that the accretion rate may drop rapidly in the absence of a mass supply. For the observed active AGN, the mass supply rate to the center of the galaxy appears to be already reduced but different from zero.

{Our comparison shows that a one-to-one match of models and observations is non-trivial. The reasons are both the limited information in observational data, but also the large parameter space in the models, as well as their time evolution. Considering the case of NGC~1097 (see Fig.~\ref{fig:NGC1097_sfr}), the overall behavior of the star formation rate is reproduced both with the models of \citet{Elmegreen2010} and \citet{Kawakatu2008}. However, we also note that significant fluctuations exist and for both models, some annuli deviate from the theoretical expectation. The latter may indicate that the star formation law is more complex and the efficiency may change depending on the environmental conditions.}

{To further improve on understanding the relation between star formation and black hole accretion, a first important step would be a more solid understanding of the star formation law. In this respect, \citet{Shetty13} recently discussed evidence for a non-universal Kennicutt-Schmidt relation, while a potential explanation for such non-universality was given by \citet{Federrath13} based on the properties of supersonic turbulence. In addition, the structure of the disk can be probed in further detail with ALMA, yielding high-resolution data on surface densities and gas kinematics. Such data, together with an indication of the current star formation rate, would strongly constrain the current state of the disk and thus provide a strong constraint on potential accretion models.}

\section{Conclusions}

In this paper, we {present a semi-analytic model based on the previous work of \citet{Kawakatu2008, Kawakatu09} describing the evolution of a black hole centered in a self-gravitating gaseous disk, that is time-dependently supplied with matter from the hosting galaxy. We have extended their model by separately considering the inner and outer disk and by considering the impact of a non-linear relation between the star formation rate, the gas surface density and the turbulent velocity.} The aim of the model is to investigate the importance of star formation in the disk for the accretion process, i.e. how the mass of the SMBH depends on the way star formation is modelled. The accretion process is described via viscous accretion in the thin disk approximation, where angular momentum is transported by kinetic viscosity. The main source of the viscosity is supersonic turbulence which is generated by stellar feedback. There we limited our model to the contribution of supernovae, as \citet{Cen2012} found, that AGN-feedback is less important than stellar feedback. Although \citet{Davies2012} state, that slow stellar winds as produced by AGB play an important role for the accretion process, we find that compared to supernovae stellar winds from OB stars as well as AGB stars inject negligable amounts of turbulent energy. Therefore we do not consider the contribution of stellar winds in our model. {We note that, even when solving only the equations given in \citet{Kawakatu2008}, our accretion rates during the initial $10^8$ yrs is different by about one order of magnitude. Here this discrepancy could not be resolved in full. Ideally, this point should thus be addressed through an independent calculation for instance based on a more detailed 3D modeling.}

Our model is able to grow SMBHs with masses of up to $10^7 M_\sun$ with mass supply rates of $1 M_\sun \mathrm{yr^{-1}}$. 
These results support the claims of \citet{Xivry2011} and \citet{Schawinski2011}, that (low-mass) SMBHs can form from accretion only. Given enough gas supplied to the nuclear region, our model is able to produce SMBHs of $10^9~M_\sun$ on similar time scales. Such large SMBHs have been observed at redshifts $z \simeq 6$ \citep[e.g.][]{Fan2001} and $z \simeq 7$ \citep{Mortlock2011}, approximately one billion years and 800 million years after the big bang respectively. This means that even these objects could have formed from accretion only, although our results do not exclude the occurence of major merger events during the evolution of the hosting galaxy, which might even be necessary to fuel the nuclear region with enough gas.

In our model, we consider the accretion process on small scales (several tens of parsecs), as the influence of large scale structures and dynamics is highly debated in the literature. For instance \citet{Nayakshin2012} conclude that SMBHs are not fed by large scale (100 pc to several kpc) gaseous structures such as disks or bars. Also \citet{Jahnke2011} explain the observed $M_\mathrm{BH}$-$M_\mathrm{bulge}$ scaling relation by a hierarchical assembly of black hole and stellar mass rather than some kind of large-scale co-evolution of SMBH and the hosting galaxy. \citet{DiamondStanic2012} found observational evidence for a strong correlation between black hole growth and star formation on smaller scales ($r<1\mathrm{kpc}$) whereas the correlation on larger scales seems to be weak. 
Nevertheless, in order to form a SMBH large amounts of gas have to be supplied to the nuclear region even larger amounts of gas have to be avaiable in the hosting galaxy.

In agreement with \citet{Montesinos2011} we found that final SMBH masses do not depend significantly on the seed mass, i.e. our model produces the same masses for black hole seeds of $10^2$ to $10^5~M_\sun$ where for seed masses of $10^6~M_\sun$ the amount of accreted matter is slightly reduced. As higher seed masses result into larger disks the gas surface density is considerably lower as the same amount of matter is spread over a larger area. This leads to significantly smaller accretion rates until the total mass of gas and stars has reached the mass of the central black hole. This effect is only visible for very large seed masses as for smaller seeds the relevant time scales are negligible compared to the total time scale. The latter implies that the growth of black holes of any mass is strongly regulated by the ambient surface densities, consistent with the findings of \citet{Shin2012}. To obtain such high surface densities at early times and close to the black hole, it is however important that the gas can collapse without efficient fragmentation \citep{Latif13a, Latif13b}. It is thus conceivable that the formation mechanism and the ambient densities are closely linked, thus providing both massive seeds and intense accretion. 

Apart from the gas surface density the turbulent velocity of the gas plays a very important role for the accretion process. As in our model turbulence is mainly driven by stellar feedback we do obtain high accretion rates only if stars are formed in the nuclear disk. The model therefore indicates a close connection between AGN activity and star formation as observed for several objects and discussed in the literature \citep[e.g.][]{Hopkins2012, Santini2012, Kumar2010, Davies2007}. 

We also stress that the choice of the star formation model, and in particular the self-regulation via turbulent velocities, has a strong impact on our results. Understanding the impact of turbulence on the star formation process in turbulent, self-gravitating disks is thus of central importance to understand what regulates the accretion of supermassive black holes, and may in fact determine the link between the  star formation rates and black hole accretion rates that was derived in observational studies. While the study of \citet{Hsieh2011} already provided high-resolution data for the giant molecular cloud complexes in NGC~1097, similar studies are required for a larger number of active galaxies under different conditions, to probe star formation on different mass scales and in environments with different degrees of rotation and turbulence. The latter may allow to break the current degeneracy in the data and to derive a star formation law that can be applied in models of the circumnuclear disks. We note in particular that a Bayesian approach, as applied by \citet{Shetty12a, Shetty13} to different environments, may help to reliably constrain such star formation models.

\appendix

\section{Average lifetime of massive stars} \label{appendix:imf}

To calculate the energy output by supernovae, we need to estimate the average lifetime of the relevant stellar population, i.e. of stars with masses between 8 and 150 solar masses.\\

As is known from stellar evolution theory, a star spends most of its lifetime on the main sequence, for which the relations between mass $M$, luminosity $L$ and lifetime $\tau$ are well known:

\begin{eqnarray}
	\tau & \propto & \frac{M}{L}	\\
	L & \propto & M^{3.5}			\\
	\tau & \propto & M^{-2.5}
\end{eqnarray}

For the sun, mass and luminosity are well known, from which the lifetime on the main sequence has been deduced as $\tau_\sun = 10^{10} \ \rm{yr}$. With this one obtains for a star of known mass the relation
\begin{equation}
	\frac{\tau}{\tau_\sun} = \left( \frac{M}{M_\sun} \right)^{-2.5} .
\end{equation}

To estimate the average lifetime of a given population we integrate the product of lifetime and relative abundance over the relevant mass range:
\begin{equation}
	\bar{\tau} = \int_{stars} \tau(M) \cdot N(M) \mathrm{d}M
\end{equation}
where $N(M) dM$ is the initial mass function.
For the relevant stars we obtain  \mbox{$\bar{\tau} = 8.7 \times 10^5~\rm{yr} \  \approx 10^6~\rm{yr}$}.

The supernova rate per solar mass of formed stars is the fraction of the stellar population bound to explode into supernovae (i.e. the population of stars with $M_* \geq 8 M_\sun$) over the total stellar population.
\begin{equation}
	f_\mathrm{SN} = \frac{\int_8^{150} N(M)~\mathrm{d}M}{\int_0^{150}N(M)~\mathrm{d}M} \approx  7.9  \times 10^{-3}
\end{equation}


\begin{acknowledgements}
 We thank W. Schmidt for fruitful discussions on the topic. D.S. and S.W. thank for financial support by the SFB~963 ''Astrophysical Flow Instabilities and Turbulence'' (project A12) via the German Science Foundation (DFG). T.~S.~P. is grateful for financial support by the RISE program of the German Academic Exchange Service (DAAD). This work was partly inspired by the COST conference \emph{Black Holes: From Quantum To Gravity} (April 2012). DRGS further acknowledges funding from the {\em Deutsche Forschungsgemeinschaft} (DFG) in the {\em Schwerpunktprogramm} SPP 1573 ``Physics of the Interstellar Medium'' under grant SCHL 1964/1-1. { We thank the anonymous referee for valuable comments on our manuscript.}
\end{acknowledgements}


\end{document}